\documentclass[12pt]{article}
\usepackage{amsmath, amsfonts, graphicx, graphics, bm, float, adjustbox, lscape, threeparttable, rotating, makecell, setspace, tabularx, multirow, booktabs, natbib, caption, subcaption, chngcntr, dsfont}
\usepackage[table,xcdraw]{xcolor}
\usepackage[section]{placeins}
\usepackage{tikz}
\usetikzlibrary{positioning}
\usepackage[utf8]{inputenc}
\usepackage[top=3cm,left=3cm,right=3cm,bottom=3cm]{geometry}
\usepackage[graphicx]{realboxes}
\usepackage [english]{babel}
\usepackage [autostyle, english = american]{csquotes}
\usepackage[all,cmtip]{xy} 
\MakeOuterQuote{"}
\usepackage[toc,page]{appendix}
\usepackage{authblk}
\usepackage[T1]{fontenc}
\usepackage{hyperref}
\hypersetup{
colorlinks=true,
citecolor=blue,
linkcolor=blue,
filecolor=blue,      
urlcolor=blue,
}

\graphicspath{{pictures}}

\begin{document}

\begin{titlepage}
\title{\Large Seeing Through Green: Text-Based Classification of Green Patents
 \thanks{\scriptsize{Authors claim funding from the European Union - NEXT-GENERATION EU, as part of the project "Fostering Open Science in Social Science Research (FOSSR)" - MUR Code PNRR-IR0000008 - CUP B83C22003950001. We would like to thank Maurizio Zanardi for his insightful suggestions.}}}

\author[a,c]{Lapo Santarlasci\thanks{\scriptsize{Email: lapo@stanford.edu. CRediT: Conceptualization, Methodology, Software, Data curation, Formal analysis, Investigation, Visualization, Writing – original draft, Resources, Validation.}}}
\author[a]{Armando Rungi\thanks{\scriptsize{Email: armando.rungi@imtlucca.it. CRediT: Writing – original draft, Funding acquisition, Conceptualization, Investigation, Methodology, Supervision, Resources, Validation.}}}
\author[b]{Antonio Zinilli\thanks{\scriptsize{Corresponding author. Email: antonio.zinilli@cnr.it. CRediT: Conceptualization, Methodology, Software, Formal analysis, Investigation, Visualization, Writing – original draft, Resources, Validation.}}}

\affil[a]{\footnotesize Firms Markets Technologies Lab, IMT School for Advanced Studies Lucca, Piazza San Francesco 19, 55100 Lucca, Italy}
\affil[b]{\footnotesize National Research Council of Italy (CNR), Research Institute on Sustainable Economic Growth, Unit of Rome, Piazzale Aldo Moro 7, 00185 Rome, Italy.}
\affil[c]{\footnotesize Stanford Institute for Human-Centered Artificial Intelligence (HAI), Stanford University, 450 Jane Stanford Way, Stanford, CA 94305, United States.}
\date{\small \today \vspace{-2.5em}}
\maketitle
\begin{abstract}
\footnotesize
This paper introduces a Natural Language Processing (NLP) approach to identify green patents from official supporting documents for the universe of about 37.8 million patent families granted in 2010-2022. We start by training a simple neural network on about 12.4 million patent applications previously classified as green in the literature. The model expands an initial dictionary of environmental technologies through vector representations of semantically related expressions. We find that text-verified green patents account for about 18.6\% of all patents classified as green by conventional methods. We document substantial heterogeneity across patent offices and technological classes. We validate our approach against a gold standard protocol, according to which patent texts were labeled by five independent annotators, and compare it with conventional classifications and with a zero-shot language model. Seven in ten patents classified as green by conventional methods are not green when read by humans, and two-thirds of green inventions are missed. Our dictionary nearly doubles precision, flags far fewer non-green patents, and identifies about half a million green patents that conventional methods overlook, with fully transparent rules.
\end{abstract}

\vspace{0cm}
\bigskip
\setcounter{page}{0}
\thispagestyle{empty}
\end{titlepage}
\pagebreak \newpage

\onehalfspacing

\maketitle

\section{Introduction}
Climate change is no longer a distant externality; it is a binding economic constraint that is already reshaping product markets, capital allocation and regulatory priorities worldwide. In the European Union, for example, the Green Deal, the Fit‑for‑55 package and a growing array of national carbon‑pricing schemes collectively tighten the link between firms' environmental footprint and their license to operate. Against this backdrop, \textit{green innovation} has emerged as a central lever for reconciling competitiveness with decarbonization goals. Yet a decade of empirical work has produced mixed verdicts on whether green innovation translates into superior firm performance. While several studies document positive stock‑market reactions or profitability gains, others find negligible or even negative effects, once R\&D crowd‑out and compliance costs or demand uncertainty are considered \citep{barbieri2017survey, przychodzen2015relationships}. Part of this disagreement may originate upstream of any estimation, in how green patents are identified in the first place. The paper therefore addresses two questions. First, how large are the errors of the conventional methods to classify green patents? Second, can a transparent text-based classifier reduce these errors, and at what cost in coverage? \\

\noindent In the first part of the paper, we challenge conventional classifications of green patents, i.e., widely used lists such as the OECD’s ENV-TECH and WIPO’s IPC Green Inventory, which assign \textit{green} status to entire patent subclasses, or the EPO’s \textsc{Y02/Y04S} tagging scheme. Because these schemes assign green status through technological classes, they risk conflating inventions with a genuine environmental purpose with other technologies that merely sit in an environmentally relevant class, and they cannot retrieve green inventions filed in classes that appear on no green list. Whether this matters is an empirical question, and we answer it against human judgment: seven in ten patents designated as green by these schemes are not green when read, and two-thirds of green patents are not retrieved by them. A second limitation concerns reproducibility: the EPO tagging also relies on search algorithms whose keywords and structure are not publicly disclosed, so that the classification cannot be rebuilt or audited by researchers.

In this paper, we propose the adoption of a Natural Language Processing method based on a simple neural network algorithm, which allows us to build a dictionary for environmental technologies that applies beyond technology classes. Using a so-called continuous bag-of-words representation, the algorithm employs a shallow neural network with a single hidden layer. The model learns semantic relationships from patent titles and abstracts and leverages them to expand the seed dictionary derived from large language models with synonyms. The model computes the 15 most semantically similar expressions for each seed. The result is a selection of what we consider text-verified green patents, which is substantially narrower (i.e., 18.6\%) than the one based on existing classifications.
\\

\noindent In addition to finding that conventional green patent classifications may suffer from Type I errors (false positives), when conventional technologies are misclassified due to broad technological class assignments, we identify Type II errors (false negatives) when genuine green innovations fall outside recognized environmental domains: our classifier flags about 2.8\% of the patents that the standard schemes leave out, and about half of them are green under human reading. This under-classification is consistent with a limited or biased selection of specific technological domains by the experts who define conventional classifications. Because our approach reads patent content directly, it can detect both kinds of error rather than inherit them from the classification codes. \\

\noindent The distinction also matters over time. For example, although the share of patents classified as green has risen in the most recent years, the share of text-verified green patents has not increased to a comparable extent.\\

\noindent Therefore, we measure both types of errors against human judgment. Five annotators labeled a stratified sample of 1,600 patent texts under a written definition of a green patent fixed before annotation. Against this gold standard, seven in ten patents designated as green by the conventional classifications are not actually green, and two-thirds of all green patents are not retrieved by them. Our dictionary nearly doubles the precision of the standard classifications and flags a far smaller share of non-green patents, with a recall that is lower but not statistically distinguishable from theirs; of the patents it adds outside the standard perimeter, about half are green, approximately half a million inventions that no classification scheme in use retrieves. We also evaluate a zero-shot language model applying the same written definition. \\

\noindent These results bear on the empirical literature with which we opened. A green-patent indicator that carries this much misclassification is a regressor measured with error; for example, in studies relating green patenting to firm outcomes, it may be estimating the effect of a diluted treatment. \\

\noindent Finally, we delve into the heterogeneity of green patents and report a few descriptive statistics on the text-verified patent applications and their differences, classified by conventional classifications. At the national patent office level, the two classification approaches diverge depending on where an invention is filed.\\

\noindent We find that, in absolute terms, the disagreements are dominated by a single office. China accounts for 3.93 million patent applications that are not found to be green according to the text analyses, while 1.11 million patent applications are retrieved outside of the perimeter of green conventional classifications. In relative terms, however, China is not an outlier. Applications that are not green by text-analyses make up 12.2\% of Chinese granted applications, a share close to South Korea (11.7\%) and below France (14.4\%). Japan (9.0\%), the United Kingdom (8.5\%) and Germany (8.4\%) follow. The share is lowest at offices such as Mexico (2.2\%), Canada (2.3\%) and South Africa (2.5\%), and at the European Patent Office it is only 3.9\%. Across technological fields, when we order by patent populations, the two approaches show a correlation of 87.5\%. For example, among the top 20 technologies, only 13 appear in both lists.\\

\noindent Finally, we examine citation patterns and find that green patents with text-confirmed environmental purposes receive more citations across all specifications, proving the citation premium stems from actual environmental content rather than just filing classes. Patents missed by standard schemes but identified by the new classifier also earn a citation premium comparable to those flagged solely by standard schemes. Overall, when comparing similar technological profiles, the green citation premium peaks (7.5\%) when both methods confirm the patent's environmental intent.  \\

\noindent The remainder of the paper is structured as follows. Section \ref{sec: lit} presents the related literature, highlighting our contribution and the rationale behind our research questions. Section \ref{sec: data desc} describes the dataset we constructed and provides some descriptive evidence. Section \ref{sec: nlp class} introduces our custom NLP classification procedure and its results. Section \ref{sec: validation} validates the classifications against human judgment. Section \ref{sec: descriptive} presents descriptive statistics on the two classifications across technologies and patent offices and examines the citations received by green patents. Finally, Sections \ref{sec: discussion} and  \ref{sec: conclusion} conclude.


\section{Related literature}
\label{sec: lit}
Our paper contributes to two distinct strands of literature. First, it primarily contributes to questioning the effectiveness of broad classification approaches used to investigate the environmental scope of patents, e.g., \cite{wagner2007relationship, johnstone2010renewable, rave2011determinants, ghisetti2021design, block2025green, rainville2025tracking}. 
Second, by understanding when a patent genuinely contributes to the \textit{green transition}, we can better discuss the limits and implications of \textit{greenwashing}, e.g., \cite{kammerer2009effects, berrone2017does}, when companies deceptively use intellectual property rights to persuade the public that they are environmentally friendly. Green patent is generally understood as a granted patent for an invention aimed at reducing negative environmental impacts or generating environmental benefits. The overarching purpose of green innovation — and, accordingly, of green patents — is to address environmental goals such as waste reduction, eco-efficiency, and emission abatement \citep{chirico2025does}.\\

\noindent Two primary approaches emerge in the literature for identifying green innovations from patents: the use of patent classification systems and keyword searches within patent texts. The former relies on established expert systems such as the International Patent Classification (IPC) of WIPO and the Cooperative Patent Classification (CPC) employed by the European Patent Office (EPO) and the United States Patent and Trademark Office (USPTO). Agencies such as the OECD, WIPO, and EPO provide dedicated lists of IPC or CPC codes corresponding to green technologies. A patent is deemed green if it falls under these codes — examples include WIPO’s Green Inventory, the EPO’s Y02 list for climate change mitigation technologies, and the OECD’s ENV-TECH classification, which covers a broader range of environmental innovations. IPC and CPC codes have become increasingly popular for identifying green patents because they are widely available and grounded in expert assessments \citep{block2025green}. Many studies therefore classify a patent as green if its codes align with recognized green technology lists, for instance, \cite{barbieri2020knowledge} used bibliographic patents data to locate environment-related patents based on the OECD ENV-TECH classification, while \cite{fabrizi2018green} employed the “Y02” scheme to target climate change mitigation technologies. Among others, \cite{bellucci2023venture} relied on a methodology developed by the European Commission’s Joint Research Centre (JRC) to construct global indicators of innovation in clean energy technologies, also using patent classifications. \cite{leon2017measuring} focused on patent families and international patent applications through WIPO’s PCT in green energy technologies, referencing WIPO’s IPC Green Inventory. Nonetheless, one acknowledged drawback of approaches based on classification codes is that categories can be overly broad. To address this issue, some scholars adopt more granular classifications by considering subclassification codes \citep{amore2016corporate}. Moreover, classification codes can be prone to subjective interpretation by patent applicants and examiners, potentially skewing results if green and non-green technologies are classified differently \citep{barbieri2020knowledge}. Discrepancies among classification systems imply that a patent may qualify as green under one system but not under another, and such a lack of harmonization in identifying green patents within IPC and CPC codes is well documented \citep{rainville2025tracking}.\\

\noindent A second approach involves mining the text of patents — particularly titles, abstracts, and claims — using environment-related or ecological keywords. \cite{wagner2007relationship}, for example, identified green innovations by scanning patent abstracts for ecological terms, while \cite{ghisetti2021design} developed a dictionary-based keyword approach informed by the ENV-TECH code descriptions to pinpoint green designs and trademarks. The European Union Intellectual Property Office (EUIPO) similarly applied a method that classifies over 85,000 descriptive terms (within the Nice\footnote{\href{https://www.wipo.int/en/web/classification-nice}{The Nice Classification (NCL)} is an international classification of goods and services applied for the registration of marks.} classification for goods and services) as green or non-green, as determined by expert evaluators \citep{euipo2021, block2025green}. Although text mining can capture green innovations that might be overlooked by classification codes, its effectiveness depends on carefully selected keywords. Patents that either do not include these specific terms or use them ambiguously may therefore be incorrectly included or excluded. For instance, \cite{nameroff2004adoption} found that “green chemistry” and “green chemical” did not appear in any patent titles, abstracts, or claims at the time of their study, underscoring a key limitation of relying solely on narrow terminology.

\noindent Against this background, we propose the construction of an environmental technology dictionary derived from large-scale training on the broadest set of patents ever classified as green by any existing scheme. Starting from \textit{seed} expressions suggested by Large Language Models, we employ a Word2Vec algorithm \citep{mikolov2013efficient, Mikolov2013DistributedRO} to identify combinations of words to be included in the baseline dictionary. Ultimately, this allows us to extract the subset of patents that we consider to be \textit{text-verified green}.
\\

\noindent Various tailored strategies have also been proposed. \cite{amore2016corporate}, for instance, adopted a detailed classification strategy that combined primary 3-digit USPTO codes with subclassifications to examine energy patents. \cite{ghisetti2021design} combined IP expertise with a semi-automated procedure to refine an initial set of keywords, thereby ensuring relevance for design- and product-oriented innovations under the OECD Environmental Technology classification. Other researchers have proposed merging classification-based and keyword-based strategies to increase accuracy (e.g., \citealp{wagner2007relationship, block2025green}). \cite{rainville2025tracking} addressed difficulties in categorizing patents related to the circular economy (CE), demonstrating that existing green patent classifications do not fully capture CE principles. They therefore suggested performing keyword analyses on patent abstracts to more accurately identify CE-related patents within the CPC scheme — a finding echoed by \cite{portillo2024circular}, who extended the notion of circular patents beyond waste-related or green patents to include those supporting a circular or sharing economy. \\

\noindent Overall, while classification-based methods remain prevalent for identifying green patents — owing to their accessibility and grounding in expert knowledge — debate persists regarding their limitations, as the chosen method can significantly influence both scholarly insights and policy recommendations. This underscores the importance of methodological rigor and transparency in identifying green patents. Our contribution relates to the stream of tailored, custom strategies by proposing a novel NLP-based classification method designed to mitigate misclassification issues. The main objective of this approach is to cut through the noise of falsely classified green patents and greenwashing practices, providing a more precise and granular perspective on the green innovation landscape and enabling policymakers to evaluate economic impacts with greater accuracy.


\section{Data}
\label{sec: data desc}

\noindent We source data from PATSTAT, a comprehensive repository of patent intelligence and statistics that contains bibliographical and legal event information on a global scale. In our analyses, we will mainly focus on patent families, as they encompass all patent documents linked to a single invention or innovation by the same inventors or assignees. Indeed, the same innovation effort can be granted intellectual property rights in more technologies and by several country offices. We will use the information at the application level when we want to demonstrate relevance at the technology or country office level. This approach allows us, at the same time, to prevent the inflation of patent counts that may arise from multiple applications across technological classes or jurisdictions, while taking into account the scope of intellectual property rights. The analyzed corpus comprises 37,797,529 patent families granted since 2010, with applications first published by 2022.

\noindent The key variables extracted from PATSTAT include the abstract and title, which provide the text summarizing the invention. We will use this text to classify and identify green patents. Additionally, we retain classification codes — specifically, the International Patent Classification (IPC) and the Cooperative Patent Classification (CPC) — that have been widely used in prior literature to identify green technologies. We also include the filing, publication, and granting dates, which together with the previous information provide a complete timeline of the innovation processes we investigate. These correspond, respectively, to the submission of a patent application to the patent office, the publication of the application, and its acceptance by the patent office. We consider the filing date the closest available proxy for the actual date of invention because it corresponds to the first time the invention is fully disclosed.

\noindent To initiate the classification process, we employ three widely recognized algorithms to identify green patents from the original PATSTAT data: the ENV-TECH, developed by the Organisation for Economic Co-operation and Development (OECD, \cite{havsvcivc2015measuring}); the EPO’s “Y02/Y04S tagging scheme” \citep{veefkind2012new}; and the WIPO–UNFCCC “IPC Green Inventory”\footnote{\href{https://www.wipo.int/classifications/ipc/green-inventory/home}{https://www.wipo.int/classifications/ipc/green-inventory/home}}. The first two are based on the International Patent Classification (IPC) and the Cooperative Patent Classification (CPC), which assign patents to specific technological fields. Table \ref{tab: ipc_envtech} provides an example of IPC codes that the ENV-TECH algorithm links to green patents.\\

\begin{table}[htpb]
    \centering
        \caption{Sample of IPC codes used by the ENV-TECH algorithm}
    \label{tab: ipc_envtech}
\resizebox{.65\textwidth}{!}{%
    \begin{tabular}{cl}
    \hline
    \textbf{IPC} & \textbf{Description}\\
    \hline
        B01D 53/34--72 & Chemical or biological purification of waste gases \\
    F01N 3 & Exhaust gas treatment in engines \\
    C02F & Treatment of water and waste water \\
    B09B & Disposal of solid waste \\
    B29B 17 & Recovery of plastics from waste \\
    \hline \hline
    \end{tabular}}
\begin{tablenotes}
\footnotesize 
\item Note: The table reports examples of IPC classes used by the ENV-TECH algorithm for the classification of patents as "green".
\end{tablenotes}
\end{table}

\noindent While the EPO Y02/Y04S tagging scheme also leverages automated text-based algorithms for green patent classification, its methodology is proprietary and non-transparent. In fact, the EPO combines classification codes with expert-designed search algorithms, but the specific keywords, search terms, and algorithmic structures are not publicly disclosed. This creates a problem of transparency and reproducibility for academic research.
Although developed independently, the three methodologies are complementary since they cover non-perfectly overlapping technological domains, so we use them in combination to obtain wider coverage for our exercise. Table \ref{tab: classification_comparison} summarizes the main differences between these methodologies. \\

\begin{table}[htpb]
    \caption{Classification codes}
    \label{tab: classification_comparison}
\resizebox{.90\textwidth}{!}{%
    \begin{tabular}{lll}
    \hline
        \textbf{Developer} & \textbf{Methodology} & \textbf{Classification codes base}   \\ \hline
        OECD & ENV-TECH (OECD) & IPC and CPC codes   \\
        EPO, UNEP, ICTSD & Y02/Y04S (EPO) & CPC codes only   \\
        WIPO \& UNFCCC & IPC Green Inventory (WIPO/UNFCCC) & IPC codes only   \\
        \hline \hline
    \end{tabular}}
\begin{tablenotes}
\footnotesize
\item Note: The table reports a summary of available classifications for the identification of green patents.
\end{tablenotes}
\end{table}

\section{Natural Language Processing and green patents}
\label{sec: nlp class}

\noindent Although classification codes are widely used to distinguish green patents, they can also lead to cases of misclassification. For instance, the code F01N7 (“Exhaust or silencing apparatus, or parts thereof”) from the ENV-TECH list is clearly associated with a technological field in which innovation often results in less polluting engines through more advanced exhaust systems. However, the same code may also include inventions aimed at enhancing engine performance by increasing fuel combustion, regardless of environmental considerations. \\

\noindent Motivated by previous evidence and several other cases of misclassification, we propose a transparent and reproducible strategy for the identification of \textit{text-verified green} patents based on the Natural Language Processing of abstracts and titles. We construct a dictionary of green technologies using a basic neural network model, following the Word2Vec algorithm \citep{mikolov2013efficient, Mikolov2013DistributedRO}, which captures semantic relationships between words. We start by extracting from PATSTAT all patent applications classified as green by any of the classification codes proposed in prior literature (Table 2), which form the training corpus described in Section 4.3. Then, we identify a basic set of \textit{seed} expressions that serve as the foundation for expanding a baseline vocabulary for green technologies. For this first stage, we use ChatGPT and Gemini, two freely available transformer-based chatbots from OpenAI and Google, each of which was asked for basic expressions denoting environmental technologies and returned 70 expressions (see Table \ref{tab: baseline_k} for the complete list of seed phrases). Both models were used off the shelf: neither was fine-tuned, and neither was specifically trained on patent documents. We merge the two lists and remove duplicates, together with expressions whose environmental meaning is not the one a patent document would give them, therefore leaving us with 107 baseline expressions reported in Table \ref{tab: baseline_k}. Then, we proceed with a vocabulary expansion by means of Word2Vec\footnote{Please refer to Appendix A for details about hyperparameter selection} embeddings. \\

\noindent The seed expressions were elicited from proprietary conversational models, whose outputs are nondeterministic and whose underlying versions change over time; a given query need not return the same list on repeated runs, so the elicitation step cannot be replicated in the strict sense. We therefore assess the sensitivity of the baseline vocabulary to this feature of its construction by regenerating comparable lists under a documented protocol. Two publicly available models of different provenance, Qwen3-14B \citep{yang2025qwen3} and Mistral-7B-Instruct-v0.3 \citep{jiang2023mistral}, are prompted with eight formulations of the elicitation task, varying both the register in which environmental technologies are described and the constraints imposed on the answer's form. Each combination is repeated 3 times at a sampling temperature of 0.7, along with 1 deterministic run at temperature 0, for a total of 64 runs.

\noindent Of the 106 distinct expressions in Table A1 (\textit{greenhouse gas} and \textit{greenhouse gases} count as one after lemmatization), 82 (77.4\%) are returned by at least one run, either verbatim or within a longer expression containing them. The 24 that are not returned are listed in Table \ref{tab: seed_regen}; with few exceptions, they are generic terms of sustainability discourse rather than designations of technologies. We read this as evidence that the baseline list is largely recoverable from a documented elicitation protocol.\footnote{We also considered grounding the elicitation in a retrieval-augmented procedure. Retrieval requires a corpus to draw from, and for the terminology of environmental technology, the available corpora are the classification schemes this paper sets out to evaluate: the OECD's ENV-TECH list, the WIPO--UNFCCC Green Inventory, and the definitions underlying the EPO's Y02 tagging. Grounding the seeds in those documents would make the dictionary inherit the boundaries of the schemes against which it is subsequently compared. The alternative grounding, the patent corpus itself, is what the Word2Vec expansion provides.}

\subsection{The model}

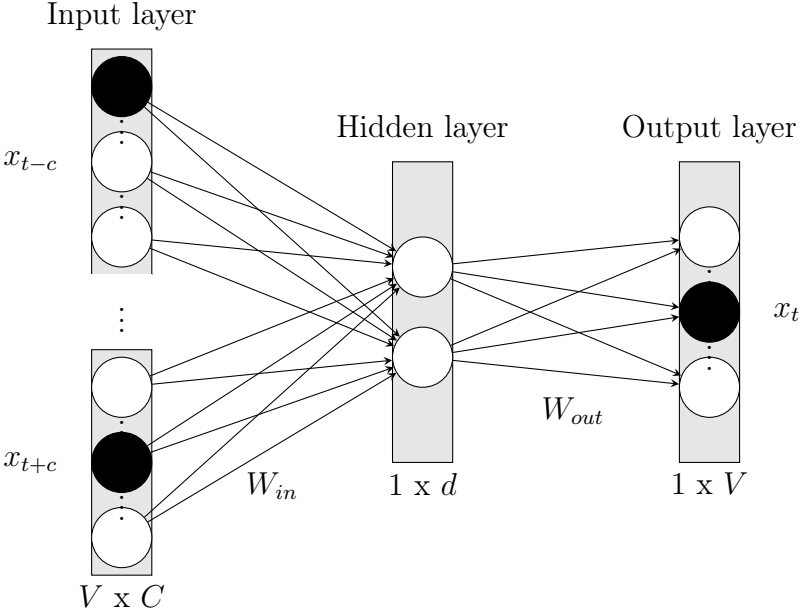
\begin{figure}[htbp]
    \begin{center}
    \caption{Neural network architecture}
    \label{fig: architecture}
\begin{tikzpicture}[
    node distance=2cm and 3cm,
    neuron/.style={circle, draw, minimum size=8mm, fill=white},
    layer/.style={rectangle, draw, minimum width=8mm, minimum height=4cm, fill=gray!20},
    >=stealth
]

\node[layer, minimum height=3cm] (input1) at (0, 6) {};
\node[above=0.1cm of input1.north] {Input layer};
\node[neuron, fill=black] (x11) at ([yshift=1cm]input1.center) {};
\node[neuron] (x12) at (input1.center) {};
\node[neuron] (x1k) at ([yshift=-1cm]input1.center) {};
\node at ([yshift=0.5cm]input1.center) {$\vdots$};
\node at ([yshift=-0.5cm]input1.center) {$\vdots$};
\node[left=0.3cm of x12] {$x_{t-c}$};

\node[layer,  minimum height=1cm, fill=white, draw=none] (inputdot) at (0, 4) {};
\node at ([yshift=0cm]inputdot.center) {$\vdots$};

\node[layer, minimum height=3cm] (input2) at (0, 2) {};
\node[below=3cm of input2.north] {$V$ x $C$};
\node[neuron] (x21) at ([yshift=1cm]input2.center) {};
\node[neuron,  fill=black] (x22) at (input2.center) {};
\node[neuron] (x2k) at ([yshift=-1cm]input2.center) {};
\node at ([yshift=0.5cm]input2.center) {$\vdots$};
\node at ([yshift=-0.5cm]input2.center) {$\vdots$};
\node[left=0.3cm of x22] {$x_{t+c}$};

\node[layer, fill=white, draw=none] (Win) at (2, 4) {};
\node[below=4cm of Win.north] {$W_{in}$};

\node[layer] (hidden) at (4, 4) {};
\node[above=0.1cm of hidden.north] {Hidden layer};
\node[below=4cm of hidden.north] {$1$ x $d$};
\node[neuron] (h1) at ([yshift=0.6cm]hidden.center) {};
\node[neuron] (h2) at ([yshift=-0.6cm]hidden.center) {};

\node[layer, fill=white, draw=none] (Win) at (6, 4) {};
\node[below=3cm of Win.north] {$W_{out}$};

\node[layer, right=of hidden] (output) {};
\node[above=0.1cm of output.north] {Output layer};
\node[below=4cm of output.north] {$1$ x $V$};
\node[neuron] (y1) at ([yshift=1cm]output.center) {};
\node[neuron,  fill=black] (y2) at (output.center) {};
\node[neuron] (yk) at ([yshift=-1cm]output.center) {};
\node at ([yshift=0.5cm]output.center) {$\vdots$};
\node at ([yshift=-0.5cm]output.center) {$\vdots$};
\node[right=0.3cm of y2] {$x_{t}$};

\foreach \i in {1,2,k} {
    \foreach \j in {1,2} {
        \draw[->] (x1\i) -- (h\j);
        \draw[->] (x2\i) -- (h\j);
    }
}
\foreach \i in {1,2} {
    \foreach \j in {1,2,k} {
        \draw[->] (h\i) -- (y\j);
    }
}
\end{tikzpicture}
    \end{center}
    \begin{tablenotes}
\footnotesize
\item Note: The figure is a representation of the basic neural network for the continuous-bags-of-words (CBOW) variant of the Word2vec algorithm.
\end{tablenotes}
\end{figure}

\noindent In Figure \ref{fig: architecture}, we represent the architecture of the bags-of-words variant of Word2vec that we use. The objective of the algorithm is to maximize the likelihood of predicting a target word given its surrounding context. Accordingly, the model computes a probability distribution over the entire vocabulary and updates the weights to minimize the prediction error between the actual target word and the predicted output. The resulting scores correspond to the likelihood of each word in the vocabulary being the correct target given the context. Notably, the bag-of-words variant of the algorithm is trained using backpropagation and stochastic gradient descent to iteratively adjust the weights. Once trained, the weights between the input and hidden layers (or the hidden layer itself) serve as word embeddings, capturing semantic relationships between words based on their co-occurrence patterns in the training corpus.

\noindent In notation let $V$ be the vocabulary size and $d$ be the embedding dimension, which we set at $450$. We define: $W_{in}$ ${\in}$ ${\mathds{R}}^{V \times d}$ as the input weight matrix and $W_{out}$ ${\in}$ ${\mathds{R}}^{d \times V}$ as the output weight matrix. For a target word $w_t$ at position $t$ with context window size $C$ we define context words $\{ {w_{t-C}, \ldots, w_{t-1}, w_{t+1}, \ldots, w_{t+C}} \}$. As from our hyperparameters' fine-tuning described in Appendix~A, we set $C=2$. For each context word ${w_{t-C}, \ldots, w_{t-1}, w_{t+1}, \ldots, w_{t+C}}$, we get their vectors:

\begin{equation}
\mathbf{x}_{t+j} \in {0,1}^V \quad \text{for } j \in {-C, \ldots, -1, 1, \ldots, C}
\end{equation}

Thus, we compute the average of word embeddings as:
\begin{align}
\mathbf{h} &= \frac{1}{2C} \sum_{j=-C, j \neq 0}^{C} \mathbf{W}_{in} \mathbf{x}_{t+j} = \frac{1}{2C} \sum_{j=-C, j \neq 0}^{C} \mathbf{v}_{w_{t+j}}
\end{align}
where $\mathbf{v}_{w_{t+j}}$ is the embedding vector for word $w_{t+j}$.

Therefore, we can use \( \mathbf{h} \) to predict the target word \( w_t \) looking at the largest value of the output vector:
\begin{equation}
\mathbf{x_t} = \mathbf{W_{out}}\mathbf{h}
\end{equation}

\subsection{Training}

\noindent The training process consists of minimizing the prediction error for the contextual word forecast. Such a process ends up tuning word embeddings so that they represent semantic relationships given the context. After training Word2Vec, each word $w \in V$ has a learned embedding vector $\mathbf{v}_w$ that we can leverage to produce synonyms for each seed expression of our baseline vocabulary. Specifically, we extract the top 15 most semantically similar words\footnote{The candidate expressions produced at this stage are screened before the vocabulary is fixed, as described in the following paragraphs.} by means of cosine similarity. In notation, for two words $w_1$ and $w_2$ embedded in vectors $\mathbf{v}_{w_1}$ and $\mathbf{v}_{w_2}$, respectively, cosine similarity is:
\begin{equation}
\text{sim}(w_1, w_2) = \frac{\mathbf{v}_{w_1} \cdot \mathbf{v}_{w_2}}{|\mathbf{v}_{w_1}| \cdot |\mathbf{v}_{w_2}|} = \frac{\mathbf{v}_{w_1}^T \mathbf{v}_{w_2}}{\|\mathbf{v}_{w_1}\|_2 \|\mathbf{v}_{w_2}\|_2}
\end{equation}
where $\|\mathbf{v}\|_2 = \sqrt{\sum_{i=1}^{d} v_i^2}$ is the Euclidean norm. \\

\noindent We apply the expanded list to the entire corpus of patents granted since 2010. We read the matched documents with attention to the cases that looked anomalous, and those readings feed back into the vocabulary: expressions that retrieve documents bearing no environmental content are removed, or made conditional on a co-occurring term, and some of the initial seeds are revised when the documents they retrieve show to be too generic for patent prose. We repeat the procedure until we converge to a stable list. Therefore, the vocabulary we report in Appendix Tables \ref{tab: baseline_k}, \ref{tab: voc_expansion_1} and \ref{tab: voc_expansion_2} is the fixed point of a loop, not the first output after the expansion. This is the reason why the vocabulary contains expressions of two kinds: single expressions, whose occurrence in a text is sufficient, and co-occurrence rules, which require a partner term within a window of twenty words. The co-occurrence rules are the record of expressions that proved too broad on their own.

\noindent Finally, we extract what we call \textit{text-verified green} patents. Importantly, we also introduce a gold-standard exercise for expert validation in Section~\ref{sec: validation}, in which five annotators, blind to any classifications, use a codebook frozen before annotation.

\noindent A natural objection to our approach is that a contextual language model, which represents words through the sentences in which they occur, would measure similarity better than a static embedding. We tested this expectation using BERT for Patents \citep{srebrovic2020bert}, with the same corpus, the same $k=15$ and the same seed expressions, varying only the model used to measure similarity. Manual screening is excluded from both pipelines, so the comparison is between the two models rather than between a model and a screening step. Both expansions are evaluated against the gold standard in Section~\ref{sec: validation} using the same matching code.

\begin{table}[!ht]
\begin{center}
\caption{Vocabulary expansion with two similarity models, before manual screening}
\label{tab: bert_vs_w2v}
\begin{tabular}{@{}lcccc@{}}
\hline
Expansion, unscreened & Expressions & Precision & Recall & F1 \\
\hline
Word2Vec & 697 & 0.220 & 0.903 & 0.354 \\
BERT for Patents & 744 & 0.194 & 0.919 & 0.320 \\
\hline
\end{tabular}
\end{center}
\begin{tablenotes}
\footnotesize
\item {Note: Both expansions start from the same 55 seed expressions (those of Table \ref{tab: baseline_k} present in the Word2Vec vocabulary), take the top 15 nearest neighbors of each, and are applied to the same corpus. Precision, recall and F1 (the harmonic mean of precision and recall) are weighted estimates relative to the gold standard in Section~\ref{sec: validation}, computed using the same matching code for both.}
\end{tablenotes}
\end{table}

\noindent At this step, the transformer does not outperform the static embedding. Starting from the same 55 seeds, BERT for Patents returns 744 expressions against 697, but with lower precision (0.194 against 0.220) and a lower F1 (0.32 against 0.35), in exchange for 1.5 points of recall. The static embedding is the more precise of the two and the better balanced. We read this not as evidence against contextual models, but as evidence that their advantage lies elsewhere. A transformer reads a term together with the sentence around it and the expansion step cannot use that, since it asks for expressions stripped of context. It can be used to read the document itself, and Section~\ref{sec: validation} reports that a zero-shot language model can do so. \\

\subsection{The application}

\noindent Our custom vocabulary for environmental technologies (Appendix Tables \ref{tab: baseline_k}, \ref{tab: voc_expansion_1}, and \ref{tab: voc_expansion_2}) {encompasses 369 expressions of two kinds: single expressions, whose presence in a text is sufficient, and co-occurrence rules, which require a main expression and at least one partner term within a window of twenty words. It is ultimately used to perform regex-based matches on patent descriptions across the whole corpus of granted patents, and not only within the sub-sample classified as green by standard algorithms. The Word2Vec model is trained on all the patent families classified as green by the three schemes, granted or not and with no restriction of year: 15.56 million patent families, of which 12.39 million with an available text enter the training, so that the embedding reflects the language in which environmental technology is written. The training corpus is deliberately broader than the corpus analyzed: more text gives the embedding more material from which to learn the vocabulary of the domain, and the vocabulary that results from it is then applied to each of the 37{,}797{,}529 patents granted since 2010.} This distinction governs what the exercise is able to show. Applied only inside the perimeter of the existing schemes, a text-based classifier can at most prune that perimeter, and the only quantity it can speak to is how many of the patents those schemes flag are false positives. Applied to the whole corpus, it also flags patents whose documents it reads as describing an environmental function but whose technological classes appear on no green list, which makes the Type II error observable as well. Section~\ref{sec: validation} measures both against human judgment. We apply the method to patent titles and abstracts in English, translating the few that are not\footnote{A few patents are not written in English; these are translated using a Python-based translator that systematically relies on Google Translate.}. We then apply a series of NLP preprocessing steps: removal of stop words, part-of-speech tagging (retaining nouns, adjectives, verbs, adverbs, and proper nouns), lowercasing, lemmatization, replacement of numbers and measurement units (e.g., LxW or LxWxH) with generic tags, and removal of special characters. Finally, we concatenate titles and abstracts into a single text string for each patent. The same preprocessing is applied to the expressions of the vocabulary, so that matching is performed on lemmatized text on both sides.\\

\noindent Because the vocabulary and the matching rule are both published, the classifier can be rebuilt from the paper. We did so, implementing it independently from Appendix Tables \ref{tab: baseline_k}, \ref{tab: voc_expansion_1}, \ref{tab: voc_expansion_2} and the rule stated above: the reimplementation agrees with the production classification on 98.9\% of the corpus; 94.3\% of the patents it flags are flagged by the production classifier, which it recovers at 81.8\%; and all of the production's positive matches are triggered by expressions published in those tables. The residual differences arise from lemmatization rather than from the vocabulary.
\\

\noindent We refer to the matched patent families as \textit{text-verified green} patents, flagging out those that unambiguously meet our criteria for environmental relevance. To illustrate the effectiveness of our approach, consider the following extract from a patent abstract that was successfully matched using our custom vocabulary: \emph{"The method for producing green fuel comprises a fermentation of vegetable materials, a separation... an anaerobic digestion of said organic material suspension... producing electricity and heat by combusting this biogas, and... supplying energy to at least one step of the method..."} This part of the abstract had already been identified as green on the basis of official classifications in existing literature. Our textual matching confirms its environmental relevance by detecting key expressions such as 'green fuel', 'anaerobic digestion', 'biogas', and the reuse of electricity and heat within the process.
\\

\noindent Notably, we also identify cases where the classification codes flagged patents as green, but our textual analysis does not confirm it. For example, patents mentioning generic terms like “carbon” without contextual relevance are discarded unless accompanied by domain-specific expressions (e.g., “carbon capture and storage” or  "carbon-based materials" in unrelated industrial applications). These false positives are filtered out by enforcing co-occurrence rules and semantic proximity constraints during the vocabulary refinement and regex matching. As an illustration of such a case, consider the following extract: \emph{"The invention relates to a method for the preparation of metal or metal oxide catalysts that are supported on porous materials. The inventive method is characterized in that it comprises the following steps consisting in: impregnating activated carbon with a catalytically-active phase or with precursors of the catalytically-active phase, shaping a paste, forming structures such as honeycomb or spheres, and subjecting the product to heat treatment to eliminate the activated carbon."}. This abstract has terms such as 'catalyst' and 'honeycomb structure', but a closer manual inspection reveals that it merely describes a general process for catalyst preparation. It describes a technical process for preparing catalysts without explicitly indicating any environmental application or purpose. 
\\

\noindent Our approach also enables us to identify green innovations in technological fields that have not been considered in prior literature. These are patents that the classification codes leave out and that the text-based classifier retrieves. Consider the case of an application with CPC codes B08B3/02, B08B3/14 and B08B13/00 (related to cleaning technologies) whose abstract states it pertains to: ``\textit{Waste liquid recycling device for surface treatment cleaning ... According to the wastewater treatment device, the liquid storage tank, the sealing gasket and the floating block are matched with one another, so that wastewater can be purified, the wastewater can be reutilized, and the loss is reduced.}''. Another application, classified under A01H1/02, A01G7/06, A01G22/00, A01H1/027 and A01K49/00 (related to methods to promote plant growth), is described as: ``\textit{Eco-friendly pollinator method for small scale crops - purpose: A method for pollination in small scale seed gathering place is provided to reduce labor cost and to collect seed… An environmentally-friendly pollination of small scale crops is performed additionally using male bees...}''. {Previous extracts illustrate the directions in which a reading of the document can depart from a reading of the classification code. \noindent However, disagreement alone would not establish which of the two approaches is closer to the concept being measured. Our claim that we retrieve \textit{text-verified green} patents is supported by the validation exercise against human judgment outlined in the following paragraphs.\\

\section{Validating the classifier against human judgment}
\label{sec: validation}

\noindent A text-based classifier earns its claims only against an external benchmark. Patent classification codes are themselves an instrument rather than a ground truth, so a comparison between two code-based schemes, or between a dictionary and a code-based scheme, cannot establish which of the two is closer to the concept meant to be measured. A benchmark has to be established starting from what the environmental purpose of an invention means; the only available external standard is the judgment of readers applying that definition to the text of a patent.

\noindent This section constructs such a standard and uses it to measure the three quantities on which any claim about \textit{text-verified green} patents ultimately rests: how often a classification flags an invention that is not green, how much of the green population it retrieves, and how large that population is. We evaluate the standard schemes as in Section \ref{sec: data desc}, the text-based classifier of Section \ref{sec: nlp class}, and a zero-shot language model applying the same written definition, on one common stratified sample of 1{,}600 patent texts labeled by five independent annotators. The comparison documents what each method captures and what it leaves out, rather than declaring a single method preferable on every count.

\noindent The exercise is, to our knowledge, the first human-validated measurement of classification error for green patents at this scale, and it is what allows the central result of the paper to be stated as a measured quantity rather than as a methodological argument.

\subsection{Design of the gold standard}\label{sec: design}

\noindent As evident from the first column of Table \ref{tab: cells}, the corpus of 37{,}797{,}529 patent families partitions into four cells according to the two binary classifications: the one proposed by previous literature and ours based on text analyses. The cells are highly unbalanced: the doubly negative cell accounts for 87\% of the corpus, while the two cells in which the classifications disagree, which are the informative ones, are comparatively small. A simple random sample would have allocated few observations to the region where the question lies. We therefore draw 400 texts from each cell and weight every observation by the inverse of its inclusion probability, so that all estimates reported below refer to the population rather than to the sample.

\noindent The strata are defined on a preliminary extraction of the two classifications, in which the classical schemes assigned green status to 5{,}416{,}750 of the granted patents; in the final version used throughout the paper, they assign it to 4{,}043{,}641, the difference consisting entirely of patents moving from green to non-green. Because the population of every combination of preliminary stratum and final cell is known exactly, the estimates are post-stratified: each text is weighted by the size of its combination over the number of usable texts it contains, and bootstrap resampling remains within the original strata. Few of the texts drawn from the preliminary \textit{conventional-only} stratum belong to the patents that changed status, so that the 1.27 million patents in that combination are represented by a small number of texts; the estimates that depend on them, chiefly that of the doubly negative cell, carry the corresponding uncertainty, which the intervals reported below include.

\noindent The sample is drawn from an extraction that also contains 4{,}133{,}359 applications that are not granted or not aggregated at family level. These are excluded from the analysis, along with 242 of the sampled texts. Because the original draw was a simple random sample within each cell, the sampled texts that fall within the restricted universe remain a simple random sample of it, and the design weights are recomputed accordingly. Of the remaining texts, 1{,}353 are usable, the others having been marked out of scope by every annotator who read them or left unresolved. No unweighted aggregate is reported in this section, since under a design of this kind an unweighted figure characterises the sample alone.

\noindent The operative definition of a green patent is fixed before annotation begins and is not revised thereafter; its SHA-256 digest is recorded in the replication package, so that the version applied by the annotators can be verified against the version we publish. The definition is written independently of the vocabulary reported in the Appendix, so that the classifier is not evaluated against a standard derived from it. Five annotators received the codebook and the texts, and were not shown either classification, so that blindness follows from the design rather than from an instruction.

\noindent All five annotators first labeled a common set of 89 texts, in a calibration round intended to establish agreement before any text is assigned to a single reader. The remainder are allocated to five lots of 423 texts, each containing the same 151 shared texts interleaved with the exclusive ones, so that agreement could be re-estimated during the production round without the annotators being aware of which texts were shared.

\begin{table}[htbp]
\centering
\caption{Inter-annotator agreement}
\label{tab: agreement}
\begin{tabular}{@{}lcc@{}}
\hline
 & Calibration round & Production round \\
 & ($n = 89$) & ($n = 148$) \\
\hline
Raw agreement & 0.892 & 0.892 \\
Fleiss' $\kappa$ & 0.759 & 0.759 \\
Krippendorff's $\alpha$ & 0.760 & 0.760 \\
Gwet's AC1 & 0.804 & 0.803 \\
\hline
\end{tabular}

\parbox{0.85\textwidth}{\footnotesize \vspace{4pt} \textit{Note:} Agreement among five annotators. The calibration round covers the 89 texts labelled in common before any text was assigned to a single reader; the production round covers the texts shared by all five lots and interleaved with the exclusive ones. Raw agreement is the proportion of annotator pairs assigning the same label. Fleiss' $\kappa$ and Krippendorff's $\alpha$ correct for chance agreement under different assumptions about the marginal distribution; Gwet's AC1 is robust to the attenuation that affects both when prevalence departs markedly from one half, which is the case here.}
\end{table}

\noindent Agreement is substantial by the usual conventions \citep{landis1977measurement} in both rounds, and the four coefficients reported in Table~\ref{tab: agreement}, raw agreement, Fleiss' $\kappa$ \citep{fleiss1971measuring}, Krippendorff's $\alpha$ \citep{krippendorff2004content} and Gwet's AC1 \citep{gwet2008computing}, are mutually consistent. When one category is much more frequent than the other, as here, $\kappa$ tends to be low even when annotators agree on most texts, because it assumes a high rate of agreement by chance; Gwet's AC1 is less affected by this imbalance and is accordingly the highest of the three chance-corrected coefficients in both rounds. Agreement measures the consistency of the readers rather than a population quantity, and is therefore computed on all calibration texts, including those that fall outside the granted universe, and on the 148 shared texts of the production round that every annotator labels. Re-estimating Table~\ref{tab: validation_results} after removing, in turn, the texts read only by each annotator leaves every comparison unchanged; the analysis is included in the replication package.

\noindent The threshold for proceeding is fixed before annotation begins at $\kappa \geq 0.60$; it is exceeded in the calibration round, so that no second round of calibration is required, and it remains exceeded throughout the production round. The codebook is left unchanged for all 1{,}600 texts, and the disagreements of the calibration round are resolved in a separate adjudication log rather than by amending the definition.

\noindent A text read by a single annotator takes that label; a text read by several takes the unanimous or majority label. Across the full annotation exercise, 1{,}356 texts rest on a single judgment, 167 are unanimous, and 72 are majority decisions. Ties are not resolved by a predefined rule; instead, they are discussed by the annotators. Four texts are resolved in this way. That the majority of texts rests on a single judgment follows from the design rather than from its execution, since five annotators covering 1{,}600 texts can overlap only on a subset without abandoning coverage of the four cells. The calibration round and the texts shared within the production round constitute the direct evidence bearing on the quality of the remainder, and we report the limitation in those terms.

\subsection{Results}
\label{sec: val_results}

\noindent Confidence intervals are bootstrap percentiles over 2{,}000 replications, resampled within cells so that the stratified design is preserved. \noindent Table \ref{tab: validation_results} reports the three classifiers side by side. The zero-shot language model (Qwen3-14B; \citealp{yang2025qwen3}) receives the codebook as the definition of its task and is not exposed to any of the evaluation labels; run with open weights at temperature 0, it is repeatable. It is the best-performing of the three on both precision and recall. The text-based classifier nearly doubles the precision of the classical scheme and has a false positive rate no higher than the language model's, with a lower but not distinguishable recall from that of the classical scheme. What separates the two text-based methods is inspectability rather than performance: the dictionary can be read, contested, and revised expression by expression, whereas the decisions of a model with fourteen billion parameters cannot. The two, therefore, address different needs, and we report both.

\begin{table}[htbp]
\centering
\caption{Classification performance against human judgment}
\label{tab: validation_results}
\resizebox{\textwidth}{!}{%
\begin{tabular}{@{}lccccc@{}}
\hline
Classifier & $n$ & Precision & Recall & F1 & False positive rate \\
\hline
Classical (IPC/CPC) & 1,353 & 0.300 [0.260--0.340] & 0.334 [0.261--0.433] & 0.316 & 0.083 [0.078--0.088] \\
Text-based (\textit{text-verified green}) & 1,353 & 0.587 [0.550--0.625] & 0.273 [0.216--0.352] & 0.373 & 0.020 [0.018--0.023] \\
Zero-shot language model & 1,353 & 0.734 [0.631--0.821] & 0.725 [0.603--0.842] & 0.729 & 0.028 [0.017--0.041] \\
\hline
\end{tabular}}
\parbox{0.92\textwidth}{\footnotesize \vspace{4pt} \textit{Note:} Estimates are weighted as described in Section~\ref{sec: design}, so that they refer to the 37,797,529 patent families in the corpus rather than to the sample. Brackets report bootstrap percentile intervals over 2,000 replications, resampled within strata. Of the 1,600 sampled texts, 242 fall outside the granted universe and 1,353 of the remainder are usable. The false positive rate is the share of non-green patents that a classifier flags as green. The language model gave no answer for 26 of the 1,353 texts; these are treated as not flagged as green.}
\end{table}

\noindent The errors of the classical classification follow from Table \ref{tab: validation_results}. Its false discovery rate, one minus precision, is 0.700 [0.660--0.740]: of every ten patents it designates as green, seven are not green under the codebook. Its false positive rate is 0.083: about one non-green patent in twelve is labeled green, and because non-green patents are so numerous, these errors outnumber the correct designations by more than two to one. Its Type II error, one minus recall, is 0.666 [0.567--0.739], so that two-thirds of the green inventions in the corpus are not retrieved. The schemes therefore both over-identify and under-identify, and the two failures are of comparable magnitude. This is the central empirical claim of the paper, and it now rests on human judgment rather than on argument.

\noindent The benchmark is the union of the three schemes described in Section 3, and this bears differently on the two errors. The Type II error is a lower bound on what any single scheme would leave out, since even the most inclusive classification available fails to retrieve two-thirds of the green population. The false discovery rate is that of the union, and a single scheme applied alone would admit fewer false positives; the over-identification we report is therefore that of the classification as it is typically assembled for wide coverage, which is the use the paper addresses. Estimated prevalence of green patents is 9.59\% of the corpus, approximately 3{,}626{,}519 patents.

\noindent The two classifications are not nested, since the text-based classifier flags patents that the classical scheme does not, and the informative quantities are accordingly cell by cell.

\begin{table}[htbp]
\centering
\caption{Estimated green patents by classification cell}
\label{tab: cells}
\begin{tabular}{@{}lrrc@{}}
\hline
Cell & Population & Of which green & Share green \\
\hline
Both classifications flag & 750,255 & 492,206 & 0.656 [0.603--0.706] \\
Text-based only & 938,201 & 498,090 & 0.531 [0.480--0.581] \\
Conventional-only & 3,293,386 & 720,428 & 0.219 [0.172--0.267] \\
Neither & 32,815,687 & 1,915,796 & 0.058 [0.036--0.084] \\
\hline
\end{tabular}
\parbox{0.85\textwidth}{\footnotesize \vspace{4pt} \textit{Note:} Population counts are exact; shares are weighted estimates on the 1,353 usable texts, with bootstrap percentile intervals over 2,000 replications resampled within strata.}
\end{table}

\noindent Around 53\% of the green patents in the corpus are identified by neither classification, approximately 1.92 million. This is the largest single block, exceeding what either method retrieves correctly. It is also the least precisely estimated: it depends on the doubly negative cell, which constitutes 87\% of the corpus and carries the largest weights, so that a small number of positives found or not found among its 395 usable texts displaces the estimate considerably. We therefore report its lower bound in preference to the point estimate --- at least 1.18 million, which already amounts to a third of all green patents --- and the same dependence accounts for the width of the recall intervals in Table \ref{tab: validation_results} relative to the precision intervals.

\noindent The contribution of the text-based match outside the classical perimeter can now be stated in green patents rather than in matched documents. Of the 938{,}201 patents it adds, 53.1\% [48.0--58.1] are green under human reading: approximately 498{,}000 green inventions that no classification scheme in use retrieves. The density of that addition, 0.531, is nearly two and a half times the density of what the classical scheme contributes on its own, 0.219.

\noindent The two classifiers trade coverage against accuracy, and they do so on opposite sides of the same perimeter. Within the classical perimeter, the text-based criterion orders patents correctly: the subset it retains is 65.6\% green, the subset it discards 21.9\%, a density three times lower. The classical scheme retrieves more green patents in absolute terms, 1.21 million against 0.99 million, but it does so by admitting 2.83 million non-green patents against 0.70 million. Each additional green patent it retrieves beyond the text-based set is obtained at the cost of nearly ten non-green ones. Measured in green patents retrieved, the two are of similar size; measured in what must be accepted to retrieve them, they are not, and that bears on whether a coefficient estimated on the resulting sample identifies anything.

\noindent Requiring agreement between the dictionary and the model raises precision from 0.734 to 0.779 but lowers recall from 0.725 to 0.237; requiring agreement among all three methods raises precision to 0.820 at a recall of 0.117. Adding the classical scheme to the model leaves precision essentially unchanged at 0.732 but lowers recall to 0.259.

\subsection{The structure of the residual error}
\label{sec: val_errors}

\noindent The false positives of the classical scheme are not evenly distributed across technologies. Of the 288 annotated patents that only the classical scheme flags, 225 are not green under the codebook, and they fall in 81 technological subclasses; five of them --- separation and filtration, batteries and fuel cells, water treatment, power supply and storage, and semiconductor devices --- account for 35\% of the false positives (Table \ref{tab: fp_subclass}). The error is large even within subclasses at the core of the environmental schemes: half of the annotated water-treatment patents and four in five of the battery and fuel-cell patents do not describe an environmental technology. This is the pattern expected when green status is assigned to an entire subclass, so that every invention filed in it inherits the label whatever its document describes.
Reading the annotated abstracts shows what lies behind the label: inventions are flagged because they belong to the same subclass as an environmental technology, or because they are used in one, rather than because they deliver an environmental function themselves, such as mounting structures and foundations for solar installations, production equipment, site-safety devices and appliances that merely run on solar power.

\begin{table}[htbp]
\centering
\caption{False positives of the classical scheme by technological subclass}
\label{tab: fp_subclass}
\begin{tabular}{@{}llcc@{}}
\hline
Subclass & Description & Share of false & Not green within \\
 & & positives (\%) & the subclass (\%) \\
\hline
B01D & Separation and filtration & 11.1 & 62.5 \\
H01M & Batteries and fuel cells & 8.9 & 83.3 \\
C02F & Treatment of water and waste water & 6.2 & 51.9 \\
H02J & Power supply, distribution and storage & 5.3 & 85.7 \\
H01L & Semiconductor devices & 3.1 & 87.5 \\
A01G & Horticulture and forestry & 2.7 & 85.7 \\
F21V & Lighting devices & 2.7 & 100.0 \\
G06F & Electric digital data processing & 2.7 & 100.0 \\
\multicolumn{2}{@{}l}{Other 73 subclasses} & 57.3 & 82.7 \\
\hline
\multicolumn{2}{@{}l}{Total} & 100.0 & 78.1 \\
\hline
\end{tabular}
\parbox{0.92\textwidth}{\footnotesize \vspace{4pt} \textit{Note:} Annotated patents flagged by the classical scheme and not by the text-based classifier (288 usable texts, of which 225 are not green under the codebook). Each patent is assigned the four-character IPC/CPC subclass recorded in the sample; the table lists the subclasses with at least six false positives. The first column reports each subclass's share of the 225 false positives; the second, the share of the annotated patents in the subclass that are not green. Since all texts in the cell carry the same weight, the shares are unaffected by weighting. In the population, the false positives of this cell correspond to approximately 2.57 million patents.}
\end{table}

\noindent The residual error of the text-based classifier has a different origin, which is syntactic rather than lexical. Patent abstracts routinely name the domain for which an invention is intended --- \textit{``a device for use in electric vehicles''}, \textit{``applied to a photovoltaic panel''}, \textit{``relates to the technical field of electric vehicle seat production''} --- and in such constructions the expression occupies the position of a complement rather than of the subject. A match over an unordered set of expressions cannot distinguish the two, and assigns the patent to the domain that its document merely names. The expression \textit{electric vehicle} illustrates the mechanism: it matches 24 gold-standard texts, only four of which are green (weighted precision 0.162), and 22 of them are retrieved by no other expression. Because the error arises from the position of the expression rather than from the expression itself, it points to a refinement of the rule rather than of the vocabulary: conditioning the match on the grammatical role of the expression.

\subsection{Robustness of the vocabulary}
\label{sec: val_sensitivity}

\noindent Robustness of the vocabulary is assessed by re-applying the matching rule to the 1{,}353 gold-standard texts under alternative specifications. Because these exercises rely on our re-implementation of the rule, the reference values (precision 0.569, recall 0.237, F1 0.335) differ marginally from those in Table~\ref{tab: validation_results}; what matters is the deviation of each variant from that reference.

\noindent Classification rests on a small core of expressions. Five of the 369 expressions account for half of the positives retrieved, twenty-five for 90\%, and 103 for all of them. The arbitrariness inherent in compiling any list is therefore confined to a few terms that can be inspected individually; the most influential, \textit{energy saving}, accounts for about 15\% of the F1. At the same time, the vocabulary is not redundant: removing expressions at random always lowers performance, by a median of 3.3, 13.7 and 34.6 \% of the F1 when 10, 25 and 50 \% of the list are dropped, and the full vocabulary lies above the 95th percentile of every random subset.

\noindent The expansion step carries most of the retrieval. The 107 seed expressions alone reach a recall of 0.058, against 0.207 for the 262 expressions added by expansion. Evaluative expressions (\textit{eco-friendly}, \textit{carbon neutral}, and the like) are rare but highly reliable, with a precision of 0.819.

\noindent The co-occurrence constraint is the single most consequential design choice. Without it, recall rises to 0.626, but precision falls to 0.402, so that the share of non-green patents among those it flags grows from 0.431 to 0.598, and the classifier reproduces the over-identification it is designed to correct. The width of the window, by contrast, is not load-bearing: between 10 and 50 words the F1 ranges from 0.328 to 0.343, within the dispersion produced by random ablation. The 20-word window is fixed in advance and is deliberately not tuned on the gold standard.

\section{Descriptive statistics}\label{sec: descriptive}
\noindent In this section, we provide descriptive statistics on the disagreements between our approach and the classification of green patents in the previous literature. We distinguish two cases: \textit{conventional-only}, a patent family that is green according to the classification codes but is not flagged by our approach, and \textit{text-based only}, a patent family that is not green according to the classification codes but is flagged by our approach.\footnote{We refer to these cases as disagreements rather than errors, since at this stage neither classification can be taken as the truth. Section~\ref{sec: validation} measures how many patents in each case are green under human reading. \\

\noindent The total numbers can be derived from the two-way Table~\ref{tab: errors}, which considers all patent families granted since 2010. In the tables that follow, we consider patent applications rather than patent families, because a single family can be filed at several patent offices and in several technologies, and only applications allow us to show how the two classifications differ across countries and technologies. The text-based only families, at the bottom left of Table~\ref{tab: errors}, are 938,201, while the conventional-only families, at the top right, are 3,293,386. Altogether, the families flagged by our approach are 4.47\% of all patent families granted since 2010, while the classification codes flag more than twice as many (10.7\%). The conventional-only families account for about 81\% of those flagged by the classification codes, and more than half (55.6\%) of the families flagged by our approach lie outside the classification codes' perimeter.}

\begin{table}[!ht]
\caption{Agreement between classification codes and the text-based classifier, patent families granted since 2010}
\label{tab: errors}
    \centering
    \resizebox{.75\textwidth}{!}{%
    \begin{tabular}{l|cccl}
    \hline
         & text-verified green & No text-verified green & Total & ~ \\ \hline 
        ~ & ~ & ~ & ~ & ~ \\ 
        Green code & 750,255 (1.98\%) & \textbf{3,293,386 (8.7\%)} & 4,043,641 (10.7\%) & ~ \\ 
        ~ & ~ & ~ & ~ & ~ \\ 
        No green code& \textbf{938,201 (2.48\%)} & 32,815,687 (86.8\%) & 33,753,888 (89.3\%) & ~ \\ 
        ~ & ~ & ~ & ~ & ~ \\ \hline
        Total & 1,688,456 (4.47\%) & 36,109,073 (95.5\%) & 37,797,529 (100.0\%) & ~ \\ \hline
    \end{tabular}}
    \begin{tablenotes}
 \footnotesize
\item Note: Off-diagonal cells (in bold) are disagreements between the text-based classifier and the classification codes of Table~\ref{tab: classification_comparison}: \textit{text-based only} (bottom left) and \textit{conventional-only} (top right). Section~\ref{sec: validation} measures how many are green under human reading. Percentages refer to all patent families granted since 2010.
     \end{tablenotes}\label{tab: errors}

\end{table}

\color{black}

\noindent When we look at the dynamics in Figure \ref{fig: green_v_true}, according to previous literature, we observe an almost constant share of green over all patents, up to a point. The share is about 9\% each year until 2021, when it starts to rise, peaking at $\simeq$ 21\% in 2022. In contrast, when we leverage our custom definition of \textit{text-verified green} patents, the share remains almost unchanged at around 4.5\% over the entire period, except for a slight increase in 2021. \\

\begin{figure}[htbp]
    \begin{center}
    \caption{Granted \textit{green} patents (grey) and \textit{text-verified green} patents (black) over total by publication year.}
    \label{fig: green_v_true}
    \includegraphics[scale=.5]{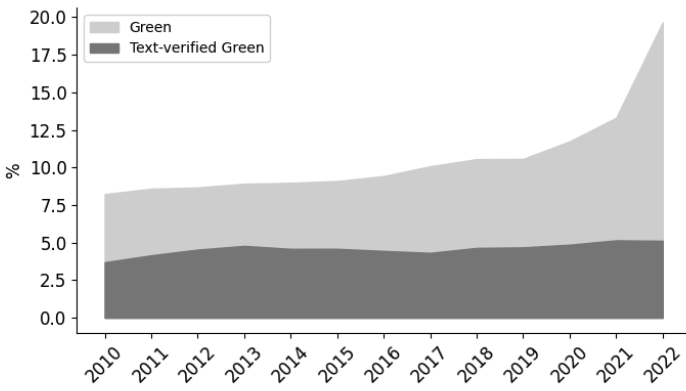}
    \end{center}
    \begin{tablenotes}
\footnotesize 
\item Note: The figure reports percentages of granted \textbf{green patents} (grey) and \textbf{text-verified green} patents (black) over total granted applications by publication year (2010-2022).
\end{tablenotes}
\end{figure}

\noindent The year 2021 has been characterized by several events likely to push for environmental innovation and practices adoptions from private firms. For example, in July the EU Commission presented the ”Fit for 55” package, a set of proposals to reduce net greenhouse gas emissions by at least 55\% by 2030, compared to 1990 levels. In April 2021 the political agreement on the European Climate Law was signed and adopted by the EU Parliament and Council in June. Innovation processes and the administrative timelines of patent filing make it unlikely that such events could be reflected in patents granted within one or two years. The share of patents flagged by the text-based classifier rises far less than the share of patents classified as green over the same years, which indicates that the recent growth of the classified-green perimeter is not matched by a proportional growth in text-verified green patents.

\subsection{Technological classes}

Table \ref{tab: tech_rank} shows how much the two classifications differ across technological fields. We report the top 20 \textit{text-verified green} technological fields by number of applications, their relevance on the total, and how their rank would change if we instead consider either existing green classifications or all patents. Although 13 technological fields out of 20 overlap between the two alternative classifications, some change rank considerably, as it is the case of "H01M - Direct conversion of chemical energy into electrical energy (e.g., batteries)", which is a top 3 according to previous literature, but it becomes 18 following our approach. At the same time, some technologies become extremely relevant under our approach, as in the case of "A61L - Methods or apparatus for sterilizing materials or objects" or "F26B - Drying solid materials or objects by removing liquid therefrom", which are 54 and 66, respectively, following previous literature. Overall, we can say that there is no correlation between the most populated technological fields considering all patents and the ones under which we find most of the \textit{text-verified green} innovations.

\begin{table}[!ht]
    \centering
    \caption{Top 20 text-verified green technologies and their rank}
        \resizebox{.99\textwidth}{!}{%
    \begin{tabular}{llccccc}
    \hline
 &&&&&Rank& \\   
        Code & Technology name & N. & \% total & text-verified green & Green Codes & All Patents \\ \hline
        B01D & Separation (e.g., filtering, distilling, centrifuging) & 131,581 & 4.76 & 1 & 1 & 6 \\ 
        C02F & Treatment of water, waste water, sewage, or sludge & 118,500 & 4.29 & 2 & 2 & 19 \\ 
        H02S & Generation of electric power by conversion of infrared, visible or ultraviolet light & 87,592 & 3.17 & 3 & 9 & 153 \\ 
        H02J & Circuit arrangements or systems for supplying or distributing electric power & 69,647 & 2.52 & 4 & 4 & 25 \\ 
        H01L & Semiconductor devices & 59,406 & 2.15 & 5 & 5 & 2 \\ 
        F24F & Air-conditioning; Air-humidification; Ventilation & 47,377 & 1.71 & 6 & 12 & 40 \\ 
        F21V & Functional features or details of lighting devices & 45,465 & 1.65 & 7 & 14 & 17 \\ 
        A01G & Horticulture & 45,113 & 1.63 & 8 & 6 & 36 \\ 
        B08B & Cleaning & 43,432 & 1.57 & 9 & 15 & 21 \\ 
        B01J & Chemical or physical processes, e.g. catalysis & 38,238 & 1.38 & 10 & 11 & 32 \\ 
        B60L & Propulsion of electrically-propelled vehicles & 37,946 & 1.37 & 11 & 8 & 115 \\ 
        A61L & Methods or apparatus for sterilising materials or objects & 36,393 & 1.32 & 12 & 54 & 54 \\ 
        F21S & Non-electric lighting devices or systems & 36,035 & 1.3 & 13 & 25 & 43 \\ 
        F24S & Solar heat collectors & 28,210 & 1.02 & 14 & 35 & 267 \\ 
        H02K & Dynamo-electric machines & 27,685 & 1 & 15 & 51 & 38 \\ 
        G01N & Investigating or analysing materials & 27,279 & 0.99 & 16 & 19 & 3 \\ 
        F21Y & Indexing scheme for lighting devices & 24,852 & 0.9 & 17 & 43 & 56 \\ 
        H01M & Direct conversion of chemical energy into electrical energy (e.g., batteries) & 24,030 & 0.87 & 18 & 3 & 16 \\ 
        F26B & Drying solid materials or objects by removing liquid therefrom & 23,780 & 0.86 & 19 & 66 & 76 \\ 
        B02C & Crushing, pulverising or disintegrating & 22,674 & 0.82 & 20 & 45 & 66 \\ \hline
    \end{tabular}}
    \label{tab: tech_rank}
    \begin{tablenotes}
 \footnotesize
     \item Note: The table shows the top 20 text-verified green technologies by number of publications and how they rank, alternatively, when we use any available green classification and the total population since 2010.
\end{tablenotes}

\end{table}

\noindent A full comparison for technological fields is reported in Appendix Tables \ref{tab: pat_dist_1}, \ref{tab: pat_dist_2}, and \ref{tab: pat_dist_3}. Figure \ref{fig: density_cpc} summarizes the share and density distribution of \textit{text-verified greens} across CPC codes. On average, the share is about 18\%, but with a dispersion that runs from 0 to 100\%\footnote{It is important to remind that a single invention effort can be acknowledged as a patent in more than one technological class, hence descriptive statistics here include multiple counting.}.

\begin{figure}[htbp]
    \begin{center}
    \caption{Kernel-density plot of the shares of \textit{text-verified green} patents by CPC category.}
    \label{fig: density_cpc}
    \includegraphics[scale=.45]{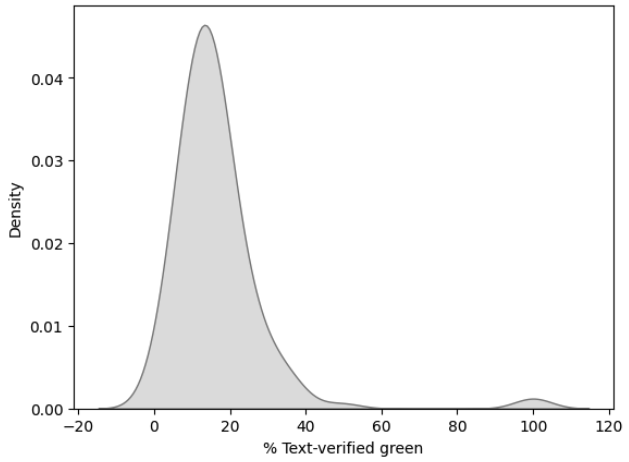}
    \end{center}
    \begin{tablenotes}
\footnotesize 
\item Note: The figure represents the kernel density of the shares of \textit{text-verified green} patents by CPC category vs. the total of green patents identified by previous classifications.
\end{tablenotes}
\end{figure}

Tables \ref{tab: tech_typeII} and \ref{tab: tech_typeI} show the top 10 technological fields for the two kinds of disagreement: applications flagged only by the classification codes, and applications flagged only by the text-based classifier. The overlap between the two tables is minimal and includes only two fields, ``B01D - Separation: Physical or chemical processes'' and ``A01G - Horticulture''. B01D is the first field in both tables, with 407,168 and 32,111 applications, respectively. Conventional-only applications are more concentrated: the first three fields in Table 10 alone account for over one million applications. In adjacent columns we show the total number of applications in each field and the share of disagreements. In ``E03F - Sewers; Cesspools'', about 84\% of the applications are flagged by the classification codes but not by the text-based classifier, and the same holds for the majority of applications in ``H01M - Direct conversion of chemical energy into electrical energy'' and ``C02F - Treatment of water, waste water, sewage, or sludge''. Text-based only applications are fewer and represent a much smaller share of each field. The two kinds of disagreement do not carry the same meaning: under human reading, about a fifth of the conventionally-only patents and about half of the text-based-only patents are green (Section~\ref{sec: validation}, Table \ref{tab: cells}).
\\

\begin{table}[!ht]
    \centering
    \caption{Top 10 technology codes: applications flagged by the classification codes only}
    \resizebox{.99\textwidth}{!}{%
     \begin{tabular}{cclccc}
    \hline
        & Code & Classification name & Conventional-only & Total patents & Share \\ \hline
        1 & B01D & Separation: Physical or chemical processes & 407,168 & 910,866 & 44.70 \\ 
        2 & H01M & Direct conversion of chemical energy into electrical energy & 304,576 & 500,123 & 60.90 \\ 
        3 & C02F & Treatment of water, waste water, sewage, or sludge & 298,459 & 463,051 & 64.45 \\ 
        4 & H02J & Circuit arrangements or systems for supplying or distributing electric power & 187,322 & 396,226 & 47.28 \\ 
        5 & H01L & Semiconductor devices; Electric solid state devices not otherwise provided for & 159,942 & 1,184,459 & 13.50 \\ 
        6 & G06F & Electric digital data processing & 108,828 & 1,920,101 & 5.67 \\ 
        7 & A01G & Horticulture & 91,535 & 314,588 & 29.10 \\ 
        8 & E04B & General building constructions & 77,117 & 300,904 & 25.63 \\ 
        9 & B60L & Propulsion of electrically-propelled vehicles & 70,522 & 161,101 & 43.78 \\ 
        10 & E03F & Sewers; Cesspools & 67,812 & 80,362 & 84.38 \\ \hline
    \end{tabular}}
    \label{tab: tech_typeII}
        \begin{tablenotes}
 \footnotesize
     \item Note: The table shows the top 10 technology codes, which show the highest numbers of applications flagged by the classification codes only, when we consider our methodology against the available green classification codes, and their shares of total patent applications.
\end{tablenotes}
\end{table}

\color{red}

\begin{table}[!ht]
    \centering
    \caption{Top 10 technology codes: applications flagged by the text-based classifier only}
        \resizebox{.99\textwidth}{!}{%

    \begin{tabular}{cclccc}
    \hline
        & Code & Classification name & Text-based only & Total patents & Share\\ \hline
        1 & B01D & Separation: Physical or chemical processes & 32,111 & 910,866 & 3.53 \\ 
        2 & A61L & Methods or apparatus for sterilising materials or objects& 27,998 & 256,482 & 10.92 \\ 
        3 & B08B & Cleaning in general; Prevention of fouling in general & 27,482 & 452,599 & 6.07 \\ 
        4 & F24F & Air-conditioning; Air-humidification; Ventilation; Use of air currents for screening & 24,336 & 296,545 & 8.21 \\ 
        5 & G01N & Investigating or analysing materials & 22,137 & 1,093,427 & 2.02 \\ 
        6 & H02K & Dynamo-electric machines & 20,571 & 300,939 & 6.84 \\ 
        7 & F21V & Functional features or details of lighting devices or systems thereof & 19,287 & 499,244 & 3.86 \\ 
        8 & A01G & Horticulture & 19,023 & 314,588 & 6.05 \\ 
        9 & F26B & Drying solid materials or objects by removing liquid therefrom & 16,660 & 208,330 & 8.00 \\ 
        10 & B01J & Chemical or physical processes & 16,600 & 358,044 & 4.64 \\ \hline
    \end{tabular}}
    \label{tab: tech_typeI}
        \label{tab: tech_typeI}
        \begin{tablenotes}
 \footnotesize
     \item Note: The table shows the top 10 technology codes, which show the highest numbers of applications flagged by the text-based classifier only, when we consider our methodology against the available green classification codes, and their shares of total patent applications.
\end{tablenotes}

\end{table}

\color{black}

\noindent The exercise we propose in this section serves a double purpose. First, we want to check which technological classes specialise more in \textit{text-verified green} patents. Second, we want to check whether our exercise identifies a different pattern of specialisation if compared with traditional classifications. To capture specialisation, we compute the so-called  \cite{balassa1965trade} index, as follows:

\begin{equation}
RCA_{c,o} \;=\; \frac{\dfrac{G_{c,o}}{N_{c,o}}}
{\dfrac{\sum_{o' \neq o} G_{c,o'}}{\sum_{o' \neq o} N_{c,o'}}},
\end{equation}

\noindent where $G_{c,o}$ is the number of text-verified green patents in technological class $c$ filed at patent office $o$, $N_{c,o}$ is the number of patents in the same class and office that are not text-verified green, and the denominator is the same ratio computed on all other patent offices $o' \neq o$ within the same class $c$. An index above 1 indicates that office $o$ is relatively more specialised in text-verified green patents in class $c$ than the rest of the world. The same expression applies to patents classified as green by conventional methods, replacing $G$ and $N$ accordingly.

\noindent We computed the RCA index for each CPC/IPC category at a 1- and 3-digit level. Figure \ref{fig: rca_density} shows the density distribution of the RCA index values by technological domains. The distribution is centered below 1 and slightly left-skewed, indicating that green patents are spread across many technological domains in which they are not concentrated.} In other words, in general, there are green patents across many technological domains, whose main scope, however, is not to produce environmental technologies. When shifting to a more granular level of classification (please refer to Appendix~B, Table \ref{tab: rca_dist_1} for top numbers of the RCA index at 3-digit of the CPC code), we observe how, specifically, top 3 positions are occupied by Greece in CPC "Semiconductor devices; electric solid-state devices not otherwise provided for" (H10), UK in "Yarns; mechanical finishing of yarns or ropes; warping or beaming" (D02), and Denmark "Saddlery and upholstery" (B68).

\begin{figure}[!ht]
    \begin{center}
    \caption{Density plot of the RCA index by CPC category.}
    \label{fig: rca_density}
    \includegraphics[scale=.5]{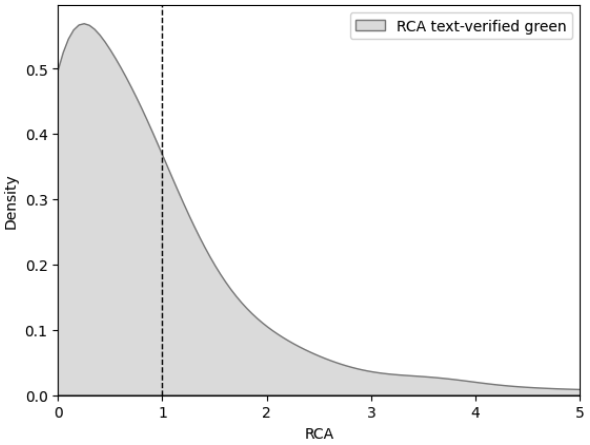}
    \end{center}
    \begin{tablenotes}
\footnotesize 
\item Note: The figure represents the density of the RCA index by CPC classification categories after applying our text-based analysis.
\end{tablenotes}
\end{figure}

\subsection{Country patent offices}
Table \ref{tab: country_errors} reports the two kinds of disagreement for the 20 patent offices with the largest number of applications since 2010. As with technological classes, the figures refer to applications rather than families, so that a single invention filed at several offices is counted at each of them. \\

\noindent The disagreements are dominated by a single office. China accounts for 3.93 million conventional-only applications and 1.11 million text-based-only applications, about 70\% and 81\% of the respective totals across the 20 offices. In relative terms, however, China is not an outlier for conventional-only applications. These make up 12.2\% of its granted applications, a share close to South Korea (11.7\%) and below France (14.4\%). Japan (9.0\%), the United Kingdom (8.5\%) and Germany (8.4\%) follow. The share is lowest at offices such as Mexico (2.2\%), Canada (2.3\%) and South Africa (2.5\%), and at the European Patent Office it is 3.9\%. This confirms that the over-identification documented in Section \ref{sec: validation} is not confined to one patent system.\\

\noindent Text-based-only green patents account for a much smaller share of each office’s applications, ranging from 0.3\% in Denmark to 3.5\% in China. The ratio between the two kinds of disagreement varies considerably across offices. For every text-based-only application, there are about 2 to 3 conventional-only applications in Brazil, Mexico and Russia, about 3.5 in China, and 8 or more in the United States, Japan, France, Austria and Denmark. Offices where the classification codes add many patents that the text does not support are therefore not the same as those where the text reveals green inventions that the codes miss. The two classifications would rank countries differently in green innovation.\\

\noindent Two cautions apply. First, a disagreement is not an error in itself. Under human reading, about a fifth of conventional-only patents and about half of text-based-only patents are green (Table \ref{tab: cells}). The gold standard is not stratified by patent office, though, so we cannot estimate office-specific error rates. Second, the text-based classifier reads English titles and abstracts. Differences across offices may partly reflect drafting conventions and the way abstracts are written or translated, as well as differences in the inventions themselves. With these qualifications, the evidence indicates that cross-country comparisons of green innovation based on classification codes depend heavily on how green patents are identified.

\begin{table}[!ht]
    \centering
    \caption{Top 20 patent offices: disagreements between the two classifications}
 \resizebox{.9\textwidth}{!}{%
    \begin{tabular}{llcccc}
    \hline
        & Country & Conventional-only & Text-based only & Share conventional-only & Share text-based only  \\ \hline
        1 & China & 3,928,409 & 1,107,390 & 12.22 & 3.45 \\ 
        2 & United States & 482,702 & 58,560 & 5.35 & 0.65 \\ 
        3 & Japan & 468,570 & 46,725 & 9.02 & 0.90 \\ 
        4 & South Korea & 339,696 & 45,795 & 11.65 & 1.57 \\ 
        5 & European Patent & 99,074 & 24,864 & 3.94 & 0.99 \\ 
        6 & Taiwan & 51,917 & 9,693 & 4.24 & 0.79 \\ 
        7 & Canada & 17,083 & 5,968 & 2.31 & 0.81 \\ 
        8 & Russia & 48,745 & 19,365 & 6.76 & 2.68 \\ 
        9 & Germany & 55,413 & 9,574 & 8.42 & 1.46 \\ 
        10 & Australia & 21,076 & 5,326 & 3.53 & 0.89 \\ 
        11 & Spain & 22,473 & 7,537 & 3.79 & 1.27 \\ 
        12 & Poland & 16,404 & 2,502 & 5.41 & 0.83 \\ 
        13 & France & 42,569 & 4,982 & 14.39 & 1.68 \\ 
        14 & Brazil & 8,567 & 4,463 & 3.00 & 1.56 \\ 
        15 & Denmark & 9,522 & 673 & 4.59 & 0.32 \\ 
        16 & Austria & 8,451 & 744 & 4.69 & 0.41 \\ 
        17 & Mexico & 3,898 & 1,693 & 2.24 & 0.97 \\ 
        18 & United Kingdom & 14,447 & 1,942 & 8.45 & 1.14 \\ 
        19 & Portugal & 3,491 & 530 & 3.04 & 0.46 \\ 
        20 & South Africa & 2,206 & 456 & 2.47 & 0.51 \\ \hline
    \end{tabular}}
    \label{tab: country_errors}
    \begin{tablenotes}
 \footnotesize
     \item Note: The table shows the 20 patent offices with the largest number of applications among patents granted since 2010, and the applications flagged only by the classification codes or only by the text-based classifier, in absolute numbers and as a share of each office's applications.
\end{tablenotes}

\end{table}

\subsection{Green citations}
In this section, we check whether \textit{text-verified green} patents are cited differently from other patents. We use citation data for patents granted between 2010 and 2022, including those classified as green by conventional methods or by our text-based classifier, as well as all patents classified as green by neither approach, for a total of 35.5 million patents in the corpus. Each patent family is represented by its first application. We first estimate the following equation on the patents classified as green by the standard schemes:

\begin{equation}
\label{eq: green citations}
    \log (1+Y_{i(j)kt})=\beta_{0}+\beta_{1}Text\_verified\_green_{i(j)}+\beta_{2}X_{i(j)}+\delta_{kt}+\gamma_{s}+\upsilon_{i(j)kt}
\end{equation}

\noindent where the dependent variable is the logarithm of the number of citations, counted at the family level, for each patent $i$ that belongs to a family $j$ in a technological class $k$ (CPC subclass), published in the earliest publication year $t$. For patents with zero citations, we add one before taking the logarithm so that they remain in the sample. The main coefficient of interest is the one on $Text\_verified\_green_{i(j)}$. 
The controls $X_{i(j)}$ include patent age and the number of family members. Finally, we add fixed effects $\delta_{kt}$ for technological class $k$ and publication year $t$, and fixed effects $\gamma_{s}$ for the cells $s$ defined by grant year, country patent office, and family size category. \\

\noindent In column 4, technology fixed effects are further interacted with the publication years. In a second step, we extend the sample to all patents and introduce three dummies: i) for patents classified as green by both approaches, ii) only by the standard schemes, iii) and only by the text-based classifier. In this case, the control group consists of patents classified as green by neither approach. Patents with no CPC class (about 2\% of the total) are excluded.\\

\noindent We report results in Table~\ref{tab: citations}. In the case of patents classified as green by the standard schemes (Panel A), those retained as \textit{text-verified} receive about 4.5\% more citations than the others, but only once fixed effects are included (columns 3 and 4). Compared with patents classified as green by neither approach (Panel B), all three groups of green patents receive more citations: about 8\% more for patents classified as green by both approaches, 4\% more for those classified as green only by our text-based classifier, and 3\% more for those classified as green only by the standard schemes (column 4). Without fixed effects (columns 1 and 2), all three groups appear to be cited much less often than non-green patents. This is a general result that reflects the characteristics of the technological classes in which patents are filed. To check the role of technological classes, we include alternative specifications in Appendix Table~\ref{tab: citations_robust}. All coefficients remain positive, and patents classified as green by both approaches remain the most cited, while the ranking between the other two groups varies across specifications.\\

\noindent We can relate to previous studies that find that green patents receive more citations than non-green ones: about 30\% more for environment-related patents \citep{barbieri2020knowledge}, and more than the other patents of the same firms for alternative energy technologies \citep{popp2012where}. Our estimates go in the same direction, with a much smaller premium. Patterns of citations also vary widely across green technologies \citep{noailly2017knowledge}, which is why it is important to compare patents within a given technological class and publication year. Two results are new. First, within the standard schemes, the patents whose environmental purpose is confirmed by the text are cited more than the others in every specification, so that the citation premium of green patents is associated with the environmental content of the inventions, not only with the classes in which they are filed. Second, the patents that the standard schemes miss but our classifier retains also receive a premium over non-green patents, comparable to that for patents classified as green only by the standard schemes. These results indicate that, once patents are compared within the same technological profile, the citation premium of green patents is small and largest for the patents whose environmental purpose is confirmed by both approaches.

\begin{table}[htbp]
\centering
\caption{Citations of green patents}
\label{tab: citations}
\small
\begin{tabular}{lcccc}
\toprule
Dep. var.: log(citations + 1) & (1) & (2) & (3) & (4) \\
\midrule
\multicolumn{5}{l}{\textit{Panel A: patents classified as green by the standard schemes}} \\
Text-verified green & $-0.046^{***}$ & $-0.024^{***}$ & $0.044^{***}$ & $0.044^{***}$ \\
 & (0.001) & (0.001) & (0.001) & (0.001) \\
Adj. R squared & 0.000 & 0.350 & 0.518 & 0.521 \\
N. obs. & 3,529,190 & 3,529,190 & 3,529,190 & 3,529,190 \\
\midrule
\multicolumn{5}{l}{\textit{Panel B: all patents, reference group: green by neither approach}} \\
Green by both approaches & $-0.391^{***}$ & $-0.097^{***}$ & $0.068^{***}$ & $0.075^{***}$ \\
 & (0.001) & (0.001) & (0.001) & (0.001) \\
Green only by standard schemes & $-0.345^{***}$ & $-0.047^{***}$ & $0.022^{***}$ & $0.027^{***}$ \\
 & (0.001) & (0.001) & (0.001) & (0.001) \\
Green only by text-based classifier & $-0.389^{***}$ & $-0.182^{***}$ & $0.040^{***}$ & $0.039^{***}$ \\
 & (0.001) & (0.001) & (0.001) & (0.001) \\
Adj. R squared & 0.010 & 0.391 & 0.633 & 0.636 \\
N. obs. & 34,668,144 & 34,668,144 & 34,668,144 & 34,668,144 \\
\midrule
Controls & No & Yes & Yes & Yes \\
CPC subclass dummies & No & No & Yes & No \\
Publication year FE & No & No & Yes & No \\
CPC subclass $\times$ publication year dummies & No & No & No & Yes \\
Grant year $\times$ office $\times$ family size FE & No & No & Yes & Yes \\
\bottomrule
\end{tabular}
\begin{minipage}{\linewidth}
\footnotesize
\textit{Note}: OLS regressions where the dependent variable is the logarithm of the family citations plus one; patents with no recorded citation are assigned zero citations. Sample: patents granted between 2010 and 2022 with available citation data; each family is represented by its first application, and each patent is one observation. CPC subclass dummies equal to 1 for each patent subclass (655 subclasses). Controls include patent age and the number of family members; patent age is absorbed by the publication year fixed effects in columns (3) and (4). Heteroskedasticity-robust standard errors in parentheses. *** p$<$0.01, ** p$<$0.05, * p$<$0.1.
\end{minipage}
\end{table}

\clearpage


\section{Discussion}
\label{sec: discussion}

This study’s primary contribution is the development and application of an NLP-based methodology for classifying  \textit{text-verified} green patents. Prior work has shown that conventional classification systems often overlook environmentally relevant patents or misclassify them, thereby failing to capture their actual environmental content \citep{block2025green}. Our findings indicate that the proposed NLP approach enhances the identification of environmentally relevant patents compared with traditional, broader classification systems. Standard methods tend to label technologies as green even when their environmental relevance is limited, inflating the count of green patents. Our validation exercise, which compares the NLP procedure with existing classifications against human judgment, shows that conventional methods substantially overestimate the number of green patents. When applying our text-analysis methodology, only about 18.6\% of patents labeled as green in the literature can be considered \textit{text-verified green}. We also find considerable heterogeneity across technological classes, with some domains more strongly aligned with genuine environmental technologies than others. \\

\noindent In addition, the analysis reveals systematic false positives and false negatives in traditional green patent classifications, which distort the understanding of environmental innovation. Measured against the gold standard, seven in ten of the patents that traditional classifications designate as green are not green, which points to a high false discovery rate, while two-thirds of all green patents are not retrieved by them; the identification of 938,201 previously overlooked patents ($\simeq 2.8\%$ during our sample period), about half of which are green under human reading, highlights the presence of significant false negatives. These inaccuracies stem from the limited scope and potential biases inherent in expert-driven selection procedures used to define classical green technology taxonomies. Our methodology nearly doubles the precision of the traditional classifications and makes their false negatives observable, with a lower but not distinguishable recall, and therefore offers more reliable indicators for policy evaluation and for directing resources toward technologies with genuine environmental impact. A zero-shot language model applying the same definition performs better on both precision and recall, which shows how much can still be gained from reading patent texts, at the cost of rules that cannot be inspected.\\

\noindent These findings have implications for research linking green patenting to firm outcomes. An indicator with this degree of misclassification is a regressor measured with error, and estimates based on it may describe the effect of a diluted treatment. The gold standard, the codebook and the code that we release allow any classification used in such studies to be evaluated against human judgment.\\

\noindent Finally, this study shows that the textual information contained in patent documents can be systematically analyzed to support a more detailed understanding of the environmental content of inventions. The proposed NLP framework is not meant to replace existing technological taxonomies but to complement them by providing a tool to assist experts in evaluating the extent to which patented technologies are genuinely environmental. Similar approaches could be adapted to other areas of technological innovation where a more fine-grained reading of patent text would help policymakers and researchers assess the implications of patenting activity more effectively.

\section{Conclusions and final remarks}
\label{sec: conclusion}

Accurately identifying green patents is essential to effective green-innovation policy. Expanding patenting incentives or promoting green-technology adoption is unlikely to deliver substantial progress if the indicators used to target and evaluate these efforts fail to distinguish environmental technologies from conventional ones.\\

\noindent Our NLP analysis finds that only about 18.6\% of patents classified as green by conventional systems are text-verified green. Human validation points in the same direction: seven in ten patents classified as green by conventional systems are not green when read by humans. This gap between classification and content suggests that broad patent classifications can lead policymakers to misallocate resources and overstate firms’ contributions to the green transition.\\

\noindent Technologies often serve multiple purposes, making it difficult to establish whether their primary purpose is environmental based on technological classes alone \citep{block2025green}. This limitation is especially relevant here: most patents identified as green by existing schemes fall within subclasses designated as environmental at the domain level. Such broad assignments may help explain why only a subset of these patents shows clear environmental content in our analysis. Moreover, major green-patent taxonomies rely on IPC and CPC structures and therefore share similar conceptual boundaries. Our NLP-based approach complements these taxonomies by assessing the text of inventions rather than relying exclusively on their assigned technology codes.\\

\noindent More accurate identification of environmentally substantive patents is also necessary to assess the economic consequences of green innovation. If conventional classifications capture only a portion of genuinely environmental inventions, analyses based on them may misrepresent the economic effects of green inventive activity. Because misclassification occurs in both directions, its effect on estimated relationships may be material. Studies of the returns to green innovation should therefore test whether their results hold under a validated classification.\\

\noindent These findings have two implications. First, green-patent indicators should be validated against an explicit definition before being used in research or policy. Second, evaluations based on conventional indicators may underestimate the economic value of genuinely environmental technologies while overstating the contribution of technologies with limited environmental relevance. By more accurately identifying green inventions, our approach supports a clearer assessment of their economic effects.\\

\noindent As policy debates increasingly call for integrated, long-term strategies for the green transition, investment in green technologies must be matched by reliable monitoring. Indicators should measure what they are intended to measure if progress toward sustainability is to be both meaningful and credible.

\section*{Data availability statement}
The data supporting the findings of this study are from PATSTAT Global (EPO). Access to these data requires a valid license, and they are not publicly available. Natural language processing (Python) code used in this study is available at \url{https://github.com/antoniozinilli/Seeing-Through-Green.git}.

\section*{AI declaration statement}
\textit{The authors acknowledge that while this work is based on their original ideas and analysis, the writing process was aided by AI tools. Specifically, Mistral's LLM was used to edit initial drafts.}
\newpage

\setlength\bibsep{0.5pt}
\footnotesize
\bibliographystyle{elsarticle-harv}
\bibliography{bibliography.bib}

\newpage


\appendix
\section*{Appendix A: Natural language processing}
\label{sec: appendix_nlp}
\setcounter{table}{0}
\renewcommand{\thetable}{A\arabic{table}}
\setcounter{figure}{0}
\renewcommand{\thefigure}{A\arabic{figure}}

\subsection*{A.1. Green-patents vocabulary}
\label{sec:vocabulary}

This section contains reference to our custom dictionary constructed for \textbf{text-verified green} patents classification. Table \ref{tab: baseline_k} contains baseline expressions related to green patents and technologies produced by ChatGPT and Gemini and used as the starting point of the vocabulary construction described in Section \ref{sec: nlp class}: the baseline was expanded with Word2Vec, and the expanded list was then screened and refined against the corpus before the vocabulary was fixed.

\begin{table}[htpb]
    \centering
        \caption{Baseline key-expressions.}
        \label{tab: baseline_k}
\resizebox{.95\textwidth}{!}{%
    \begin{tabular}{llll}
    \hline 
    \textbf{Baseline keywords} & & & \\
    \hline
        air purification & eco friendly & green design & recycled packaging \\ 
        alternative energy & eco innovation & green innovation & recycled products \\ 
        anaerobic digestion & eco solutions & green manufacturing & regenerative braking \\ 
        bio-based materials & ecodesign & green technology & renewable energy \\ 
        biodegradable polymers & ecological design & green transportation & resource efficient \\ 
        biodegradable products & ecological footprint & greenhouse gas & resource recovery \\ 
        biodiversity & electric motors & greenhouse gases & solar geoengineering \\ 
        biofuel & electric vehicles & hybrid vehicles & solar panels \\ 
        biomass & energy conservation & hydroelectric dams & sustainability metrics \\ 
        bioplastics & energy efficiency & hydroponics & sustainability practices \\ 
        carbon capture & energy efficient & hydropower & sustainable agriculture \\ 
        carbon filters & energy optimization & leed certified & sustainable development \\ 
        carbon neutral & energy recovery & life cycle & sustainable farming \\ 
        carbon offsetting & energy savings & lightweight materials & sustainable forestry \\ 
        carbon sequestration & environmental certification & low emission & sustainable manufacturing \\ 
        cellulosic plastics & environmental compliance & low impact & sustainable materials \\ 
        circular economy & environmental conservation & material compatibility & sustainable practices \\ 
        clean energy & environmental labeling & organic farming & vertical farming \\ 
        clean production & environmental management & photocatalysis & waste minimization \\ 
        clean technology & environmental policy & plant based & waste recycling \\ 
        climate adaptation & environmental stewardship & pollution control & waste reduction \\ 
        climate friendly & fuel cells & precision agriculture & water conservation \\ 
        compostable materials & geothermal & product lifespan & water efficient \\ 
        composting & geothermal power & public transportation & wind turbines \\ 
        dioxide capture & global warming & rainwater harvesting & zero emissions \\ 
        eco conscious & green buildings & recyclable materials & zero waste \\ 
        eco efficient & green chemistry & recycled materials & ~ \\ 
        \hline \hline
    \end{tabular}}
\begin{tablenotes}
\footnotesize 
\item Note: The table reports the list of baseline key-expressions related to green technologies generated by ChatGPT and Gemini. The table lists 107 entries; \textit{greenhouse gas} and \textit{greenhouse gases} coincide once the text is lemmatized, so the baseline comprises 106 distinct expressions. From this baseline we performed the vocabulary expansion described in Section \ref{sec: nlp class}.
\end{tablenotes}
\end{table}

The baseline expressions of Table \ref{tab: baseline_k}, together with those listed in Tables \ref{tab: voc_expansion_1} and \ref{tab: voc_expansion_2}, make up our final enlarged vocabulary for the regex match on patent texts. It comprises 369 expressions of two kinds:
\begin{itemize}
    \item \textbf{Single expressions} like \textit{energy conservation} that, if matched into a text, would label it as green related.
    \item \textbf{Co-occurrences} like \textit{carbon} and \textit{capture} that require the presence of both \textbf{within} a distance of 20 words to classify a patent text as green.
\end{itemize}

\begin{table}[htbp]
\renewcommand{\thetable}{A1.1}
\begin{center}

\caption{Baseline expressions not recovered by the seed regeneration}
\label{tab: seed_regen}
\small
\begin{tabular}{@{}lll@{}}
\hline
cellulosic plastics & ecological design & plant based \\
clean production & environmental certification & public transportation \\
climate friendly & environmental labeling & recycled materials \\
eco conscious & environmental stewardship & recycled packaging \\
eco efficient & hydropower & recycled products \\
eco innovation & lightweight materials & resource efficient \\
eco solutions & material compatibility & solar geoengineering \\
ecodesign & photocatalysis & water efficient \\
\hline
\end{tabular}
\end{center}
\begin{tablenotes}
\footnotesize
\item Note: The table reports the 24 expressions of Table \ref{tab: baseline_k} that none of the 64 regeneration runs returned, at any level of matching. Nineteen of the 24 are generic terms of sustainability discourse rather than designations of technologies.
\end{tablenotes}
\addtocounter{table}{-1}
\end{table}

\begin{table}[htpb]
\centering
\caption{Additional keywords for REGEX match (part I).}
\label{tab: voc_expansion_1}
\resizebox{1\textwidth}{!}{%
\begin{tabular}{ll|ll}
\hline
\textbf{Keyword} & \textbf{Co-occurrence} & \textbf{Keyword} & \textbf{Co-occurrence} \\
\hline
acidogenesis & wastewater, pollutant & disinfection & non-chlorine, ultraviolet \\
        activatedcarbon & filtration, purification, pollutants & disinfector & non-chlorine, ultraviolet \\
        adsorption & emissions \& capture & dmfc & direct methanol fuel cell \\
        aerodynamic design & - & ecofriendly & - \\
        aerogenerator & - & ecologic & - \\
        aeroponic & hydroponics, vertical farming & ecological restoration & - \\
        aeroponic cultivation & hydroponics, vertical farming & ecologically & - \\
        aeroponic culture & hydroponics, vertical farming & ecologization & - \\
        aeroponics & cultivation, cropping & ecology & - \\
        afforestation & - & ecosystem & - \\
        bamboo charcoal & sustainable forestry & ecotoxicity & avoid, reduce, reduction \\
        bioactivity & - & electric motor & - \\
        bioaugmentation & - & electric powertrain & - \\
        biocarbon & - & electric vehicle & - \\
        biochemical & biofuels, bioplastics & electricvehicle & - \\
        biochemistry & biofuels, bioplastics & electrochemical & fuel cell \\
        biocomposite & - & electrodialysis & desalination, wastewater treatment \\
        bioconversion & biomass, renewable energy & electrolyzer & hydrogen production, water electrolysis \\
        biodegradable & - & emission reduction & - \\
        biodegradable compostable & - & energy save & - \\
        biodegradable polymer & - & energy saving & - \\
        biodegradable product & - & energysaving & - \\
        biodegrade & - & energysource & clean, sustainable, renewable \\
        biodiesel & - & engine optimization & - \\
        bioenergy & - & environmental & sustainability, pollution reduction \\
        bioethanol & - & environmental impact & reduce, lower \\
        biofertilizer & - & environmental risk & reduce, lower \\
        bioleache & - & environmentally benign & - \\
        biologically degradable & - & environmentally friendly & - \\
        bioplastic & - & environmentally sustainable & - \\
        biopolymer & - & enzymolysis & biodegradable materials, biomass conversion \\
        bioproduct & - & ethanol & biofuel, renewable energy \\
        bioremediation & - & evaporative crystallization & water treatment, desalination \\
        biotechnology & - & fenton & wastewater treatment, pollutant degradation \\
        butanol & biofuels & ferment & biofuels, biogas production \\
        carbon footprint & reducing & fermentable sugar & biofuels, biogas production \\
        carbon offset & - & fertigation & - \\
        carbonfilter & - & filter & air pollution control, water treatment \\
        carwash & water-efficient & flywheel & energy storage, regenerative braking \\
        catalyzing & - & foodwaste & reduction, avoid \\
        claus & sulfur recovery, tail gas treatment & footprint accounting & - \\
        co$_2$ & capture, reduction, reduce, remove, removal & forward osmosis & - \\
        coemission & avoid, reduction, reduce, remove, removal & fuelcell & - \\
        compostable & - & geothermy & - \\
        compostable material & - & greening & - \\
        composted & - & greywater & water reuse, wastewater treatment \\
        conservancy hydropower & - & humification & composting, organic matter \\
        conservation minded & - & hybrid powertrain & - \\
        cropping practice & sustainable, organic & hybrid vehicle & - \\
        daylighting & energy saving, energy efficient & hydroelectric & - \\
        decarbonation & - & hydroelectric dam & - \\
        decarbonization & - & hydrolysed & biomass conversion, biofuel production \\
        dechlorinate & - & hydrolytic acidification & biomass conversion, biofuel production \\
        decyanation & - & hyperfiltration & water treatment, desalination \\
        deduste & air pollution & insulate & energy efficiency, building insulation \\
        demercuration & - & insulated & energy efficiency, building insulation \\
        denitride & air pollution & insulating & energy efficiency, building insulation \\
        denitrifying & water pollution & insulative & energy efficiency, building insulation \\
        denitrogenation & air pollution & insulator & energy efficiency, building insulation \\
        denox & air pollution & lignocellulose & biomass conversion, biofuel production \\
        depollute & - & lignocellulosic & biomass conversion, biofuel production \\
        depollution & - & livestock excreta & biogas production, waste management \\
        desalinate & energy-efficient & livestock manure & biogas production, waste management \\
        desalination & energy-efficient & low carbon & emission \\
        desalinization & energy-efficient & mangrove forest & - \\
        desalt & energy-efficient & membrane & filtration, water treatment \\
        desulfurate & air pollution & membrane filtration & water treatment \\
\hline \hline
\end{tabular}
}
\begin{tablenotes}
\footnotesize 
\item Note: The table reports the expressions used for the regex match on patent texts. They originate in the Word2Vec expansion of the baseline expressions of Table A1 and were then screened and refined as described in Section~\ref{sec: nlp class}.
\end{tablenotes}
\end{table}

\begin{table}[htpb]
\centering
\caption{Additional keywords for REGEX match (part II).}
\label{tab: voc_expansion_2}
\resizebox{1\textwidth}{!}{%
\begin{tabular}{ll|ll}
\hline
\textbf{Keyword} & \textbf{Co-occurrence} & \textbf{Keyword} & \textbf{Co-occurrence} \\
\hline
        membrane separation & water treatment & renewably & energy, resource, material \\ 
        methanation & biogas production, renewable energy & reprocess & waste management, resource recovery, resource, waste, material \\ 
        methanogenesis & biogas production, renewable energy & reprocessed & waste management, resource recovery, resource, waste, material \\ 
        microfiltration & water treatment, wastewater treatment & reprocessing & waste management, resource recovery, resource, waste, material \\ 
        microgrid & renewable energy, distributed generation & resource & efficiency, conservation, reuse \\ 
        nanobubble & water treatment, air purification & restoration & ecological restoration, land reclamation \\ 
        nanofilter & water treatment, wastewater treatment & reusable & material, product, resource \\ 
        nanofiltration & water treatment, wastewater treatment & reuse & material, product, resource \\ 
        natural ventilation & energy efficient, energy saving & reusing & waste, material, resource \\ 
        nitrifying denitrify & wastewater treatment, nitrogen removal & reutilization & waste, material, resource \\ 
        oilfield reinjection & enhanced oil recovery, carbon capture & reutilize & waste, material, resource \\ 
        orc & organic rankine cycle, waste heat recovery & reutilizing & waste, material, resource \\ 
        ordered mesoporous & catalysis & reverse osmosis & low energy, sustainable, reuse, recycle, low power \\ 
        oxyfuel & - & reverseosmosis & low energy, sustainable, reuse, recycle, low power \\ 
        oxygenator & aquaculture, wastewater treatment & rice husk & biomass conversion, renewable energy source \\ 
        oxygenerator & aquaculture, wastewater treatment & rooftop greening & - \\ 
        ozonator & water treatment, air purification & saccharify enzyme & biofuel production, biomass conversion \\ 
        ozone & water treatment, air purification & seawater desalination & low energy, sustainable, reuse, recycle, low power \\ 
        ozone depletion & reduce, prevent & seawater desalting & low energy, sustainable, reuse, recycle, low power \\ 
        ozone generators & water treatment & seedling raising & sustainable agriculture, forestry \\ 
        ozonize & water treatment, air purification & semicoke & biomass conversion, coal alternative \\ 
        ozonizer & water treatment, air purification & sequester & carbon dioxide, co$_2$, greenhouse gases \\ 
        ozonolysis & water treatment, air purification & sequestrate & carbon dioxide, co$_2$, greenhouse gases \\ 
        peak shaving & energy management, energy storage & sequestration & carbon dioxide, co$_2$, greenhouse gases \\ 
        pemfc & proton exchange membrane fuel cell, hydrogen fuel cell & simultaneously practice thrift & resource efficiency, waste reduction \\ 
        pervaporation & membrane separation, desalination & smart grid & renewable energy integration, energy efficiency \\ 
        photocatalyst & - & smart grids & renewable energy integration, energy efficiency \\ 
        photocatalytic & - & sncr denitration & flue gas treatment, air pollution control \\ 
        photochemistry & solar energy conversion & sofc & solid oxide fuel cell, clean energy \\ 
        photodegradable & - & soilless cultivation & - \\ 
        photolysis & - & soilless culture & - \\ 
        photolytic & - & solar & energy, cell, panel, power \\ 
        photooxidation & - & solarenergy & - \\ 
        photoreaction & solar energy conversion & solarpanel & - \\ 
        photosynthetic & biofuels & solarpower & - \\ 
        photovoltaic & - & straw & biomass resource, renewable energy \\ 
        pollutant & reduce, reduction, capture, remove, removal & sugar beet & biofuel feedstock, renewable energy \\ 
        polluted & clean, filter & sun shading & sustainable design, building efficiency \\ 
        pollution & reduce, reduction, capture, remove, removal, avoid & sunshade & sustainable design, building efficiency \\ 
        polution & reduce, reduction, capture, remove, removal, avoid & supercapacitor & energy storage, renewable energy integration \\ 
        polylactide & - & supercharge & electric vehicle, battery \\ 
        poultry farming & sustainable agriculture, waste management & supercharging & electric vehicle, battery \\ 
        poultry manure & sustainable agriculture, waste management & sustainability & goal, practice, system, production \\ 
        propeller & electric, hybrid, low-emission & sustainable & system, production, process, technology, resource, energy \\ 
        propulsion & electric, hybrid, low-emission & thermocatalytic & biomass conversion, waste valorization \\ 
        propulsor & electric, hybrid, low-emission & thermochemical & biomass conversion, renewable energy \\ 
        pyrohydrolysis & waste conversion, biomass conversion & ultrafilter & water treatment, purification \\ 
        recuperation & energy, waste, material, resource & ultrafiltrate & water treatment, purification \\ 
        recyclable & energy, waste & ultrafiltration & water treatment, purification \\ 
        recyclable material & - & ultralight & material, construction \\ 
        recyclate & energy, waste, material, resource & ultralow & emission, power consumption \\ 
        recycle & energy, waste, material, resource & urban planning & sustainable development, green infrastructure \\ 
        recycle material & - & uv photolysis & water treatment, pollutant degradation \\ 
        recycle practice thrift & - & waste management & recycling, resource recovery \\ 
        recycle product & - & waste recycle & - \\ 
        recycleable & - & wasted & recycling, resource recovery, avoid \\ 
        recycled & energy, waste, resource & wastepaper & recycling, paper recycling, avoid \\ 
        recycling & energy, waste, material, resource & water purification & - \\ 
        regasification & liquefied natural gas, renewable energy sources & water treatment & low energy, reuse, recycle, efficient \\ 
        regenerate & soil, environment, material, resource & waterpower & - \\ 
        regenerated & soil, environment, material, resource & windmill & - \\ 
        regeneration & soil, environment, material, resource & windpower & - \\ 
        regenerative & braking, energy recovery & windturbine & - \\ 
        remediation & environmental cleanup, contaminated soil & windwheel & - \\ 
        renewability & energy, waste, material, resource & woody & biomass, biofuel \\ 
        renewable & energy, resource, material & ~ & ~ \\ 
\hline \hline
\end{tabular}
}
\begin{tablenotes}
\footnotesize 
\item Note: The table reports the expressions used for the regex match on patent texts. They originate in the Word2Vec expansion of the baseline expressions of Table A1 and were then screened and refined as described in Section~\ref{sec: nlp class}.
\end{tablenotes}
\end{table}


\subsection*{A.2. Word2Vec specification}
\label{sec: appendix_finetuning}

\begin{figure}[htbp]
    \begin{center}
    \caption{Grid search for the selection of \emph{context window} and \emph{minimum count} hyperparameters.}
    \label{fig: ftune}
    \includegraphics[scale=.8]{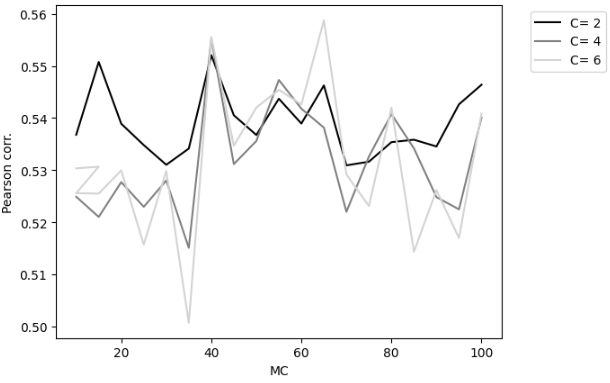}
    \end{center}
    \begin{tablenotes}
\footnotesize 
\item Note: The figure represents the hyperparameters' selection for the Word2Vec model. Each point corresponds to a combination of parameters, and the related linear correlation value with our benchmark similarity standard (the higher the better).
\end{tablenotes}
\end{figure}

In this section, we provide details about the selection of the Word2Vec model's hyperparameters. We need to find the context window size ($C$), which defines the context length for training the neural network, and the minimum count ($MC$), which sets the frequency threshold for a word to be included in the training corpus. We employ a data-driven tuning process. We use WordSim353, which is a test collection for measuring word similarity or relatedness \citep{Agirre_et_al_2013}. We then train multiple Word2Vec models using a grid of different $C$ and $MC$ parameter combinations. For each trained model, we calculated the cosine similarity for word pairs in the WordSim353 list and measured the Pearson correlation between our results and the gold standard scores. As shown in Figure \ref{fig: ftune}, all tested parameter pairs yielded a positive correlation above 50\%. Generally, models with a smaller $C$ performed better when the minimum count was held constant. This empirical process pointed towards optimal values of approximately $MC$=40 and $C$=2, which we adopt in our analysis in the manuscript.\\

\begin{figure}[htbp]
    \begin{center}
    \caption{Grid search for the selection of the embedding vector size.}
    \label{fig: ftune_embedding}
    \includegraphics[scale=.5]{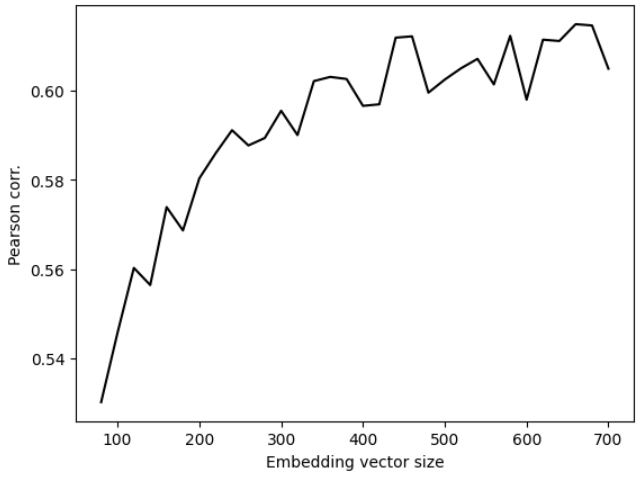}
    \end{center}
    \begin{tablenotes}
\footnotesize 
\item Note: The figure represents the selection of the embedding dimension for the Word2Vec model. Each point corresponds to a value of vector size, and the related linear correlation value with our benchmark similarity standard (the higher the better).
\end{tablenotes}
\end{figure}

\noindent We use the same procedure for fine tuning the embedding vector size $d$ (first layer of our shallow neural network). Figure \ref{fig: ftune_embedding} shows correlation with our benchmark for different sizes of embeddings. Given the convergence around 0.61 for embedding with size greater than 450, we choose such value in order to get the best empirical performances while minimizing the model complexity and overcoming limitations on computational power in larger settings.

\subsection*{A.3. Combinations of classifiers}

\noindent Section~\ref{sec: val_results} evaluates the three classifiers one at a time. Because they rest on different information --- the technological classes assigned to a patent, a published vocabulary matched against its text, and a language model reading the same text under the codebook --- they need not err on the same patents, and requiring agreement between them could in principle remove errors that each makes on its own. Table~\ref{tab: combinations} reports the performance of every such combination against the gold standard, with the same weights as Table~\ref{tab: validation_results}.

\noindent Requiring agreement raises precision but lowers recall more than it raises precision, so that no combination improves on the language model taken alone. Adding the dictionary to the model raises precision from 0.734 to 0.779 and lowers recall from 0.725 to 0.237. Adding the classical codes leaves precision essentially unchanged, at 0.732, and lowers recall to 0.259, which indicates that the codes carry no information the model does not already extract from the text. Relaxing the requirement to agreement between any two of the three methods recovers part of the recall, but at a precision below that of the model alone. The combinations therefore confirm the ordering of Table~\ref{tab: validation_results} rather than altering it.

\begin{table}[htbp]
\centering
\caption{Combinations of classifiers against human judgment}
\label{tab: combinations}
\begin{tabular}{@{}lccccc@{}}
\hline
Classifier & $n$ & Precision & Recall & F1 & False positive rate \\
\hline
Classical (IPC/CPC) & 1,353 & 0.300 & 0.334 & 0.316 & 0.083 \\
Text-based (\textit{text-verified green}) & 1,353 & 0.587 & 0.273 & 0.373 & 0.020 \\
Zero-shot language model & 1,353 & 0.734 & 0.725 & 0.729 & 0.028 \\
\hline
All three agree & 1,353 & 0.820 & 0.117 & 0.205 & 0.003 \\
Text-based and classical & 1,353 & 0.656 & 0.136 & 0.225 & 0.008 \\
Text-based and language model & 1,353 & 0.779 & 0.237 & 0.364 & 0.007 \\
Classical and language model & 1,353 & 0.732 & 0.259 & 0.383 & 0.010 \\
At least two of three & 1,353 & 0.686 & 0.398 & 0.503 & 0.019 \\
At least one of three & 1,353 & 0.443 & 0.817 & 0.574 & 0.109 \\
\hline
\end{tabular}
\parbox{0.92\textwidth}{\footnotesize \vspace{4pt} \textit{Note:} Weighted estimates on the gold standard, as in Table~\ref{tab: validation_results}. Combinations joined by ``and'' flag a patent as green only when every classifier named does; ``at least $k$ of three'' flags it when $k$ or more of the three do. The false positive rate is the share of non-green patents flagged as green. The language model abstained on 26 of the 1,353 texts; these are counted as non-green.}
\end{table}

\subsection*{A.4. Robustness of the citation estimates}

\noindent Table~\ref{tab: citations_robust} reports the specification of column (4) of Table~\ref{tab: citations} under alternative treatments of technological classes and of uncited patents. In rows (b) and (c), each pair of patent and CPC subclass is a separate observation, unweighted or weighted by the inverse of the number of subclasses of the patent, with standard errors clustered by patent. In row (d), the technological class is the alphabetically sorted combination of all the CPC subclasses of each patent (878,486 combinations, 98.4\% of patents sharing theirs with at least one other patent). Row (e) excludes the patents with no recorded citation, with the specification of row (b).

\begin{table}[htbp]
\centering
\caption{Citations of green patents: alternative specifications}
\label{tab: citations_robust}
\resizebox{\textwidth}{!}{%
\begin{tabular}{lccccc}
\toprule
 & Panel A & \multicolumn{3}{c}{Panel B} & \\
\cmidrule(lr){2-2}\cmidrule(lr){3-5}
Specification & Text-verified & Both & Standard & Text-based & N. obs. \\
 & green & approaches & schemes only & only & (Panel B) \\
\midrule
(a) Subclass dummies (Table~\ref{tab: citations}) & $0.044^{***}$ & $0.075^{***}$ & $0.027^{***}$ & $0.039^{***}$ & 34,668,144 \\
 & (0.001) & (0.001) & (0.001) & (0.001) & \\
(b) Patent $\times$ subclass observations & $0.052^{***}$ & $0.114^{***}$ & $0.061^{***}$ & $0.047^{***}$ & 53,923,237 \\
 & (0.001) & (0.001) & (0.001) & (0.001) & \\
(c) As (b), weighted & $0.052^{***}$ & $0.110^{***}$ & $0.055^{***}$ & $0.045^{***}$ & 53,923,237 \\
 & (0.001) & (0.001) & (0.001) & (0.001) & \\
(d) Combination of subclasses & $0.047^{***}$ & $0.075^{***}$ & $0.027^{***}$ & $0.042^{***}$ & 34,668,144 \\
 & (0.001) & (0.001) & (0.001) & (0.001) & \\
(e) As (b), cited patents only & $0.041^{***}$ & $0.087^{***}$ & $0.048^{***}$ & $0.042^{***}$ & 36,504,731 \\
 & (0.001) & (0.001) & (0.001) & (0.001) & \\
\bottomrule
\end{tabular}}
\begin{minipage}{\textwidth}
\vspace{4pt}
\footnotesize
\textit{Note}: Each row reports the coefficients of column (4) of Table~\ref{tab: citations} (controls, technological class $\times$ publication year and grant year $\times$ office $\times$ family size fixed effects) under the specification indicated: (a) one observation per patent with a dummy for each of its CPC subclasses, as in Table~\ref{tab: citations}; (b) one observation per patent and CPC subclass; (c) as (b), weighted by the inverse of the number of subclasses of the patent; (d) one observation per patent, with the technological class defined as the combination of its CPC subclasses; (e) as (b), excluding patents with no recorded citation. Panel B coefficients are differences with patents classified as green by neither approach. Standard errors in parentheses, clustered by patent. *** p$<$0.01, ** p$<$0.05, * p$<$0.1.
\end{minipage}
\end{table}

\section*{Appendix B: Tables and graphs}
\label{sec:appendix_tab_graph}
\setcounter{table}{0}
\renewcommand{\thetable}{B\arabic{table}}
\setcounter{figure}{0}
\renewcommand{\thefigure}{B\arabic{figure}}


\subsection*{B.1. Classification results}
\label{sec:class_res}

In this section, we provide descriptive statistics about both kinds of green patents classification implemented in this analysis, i.e. \textit{green} and \textit{text-verified green}. Specifically, Tables \ref{tab: pat_dist_1}, \ref{tab: pat_dist_2} and \ref{tab: pat_dist_3} contain number of patents, green patents and \textit{text-verified green} patents granted in every CPC/IPC technological domain (at a 3-digits level). Table \ref{tab: rca_dist_1} represent our RCA index computed for each patent office - CPC classes (3-digit level definition); top 10 couples for green and text-verified green version are showed.

\begin{landscape}
\begin{table}[htbp]
    \centering
    \caption{Patents distribution by CPC (part I).}
        \label{tab: pat_dist_1}
\resizebox{1.3\textwidth}{!}{%
            \begin{tabular}{clcccc}
            \hline
        \textbf{Classification code} & \textbf{Description} & \textbf{Patents} & \textbf{Green patents} & \textbf{text-verified green patents} & \textbf{text-verified green\%} \\ 
        \hline
        A01 & Agriculture;forestry; animal husbandry; hunting;trapping; fishing & 918,032 & 246,236 & 42,903 & 17.42 \\ 
        A21 & Baking; edible doughs & 46,668 & 1,742 & 363 & 20.84 \\ 
        A22 & Butchering; meat treatment; processing poultry or fish & 26,674 & 1,719 & 230 & 13.38 \\ 
        A23 & Foods or foodstuffs; treatment thereof, not covered by other classes & 340,445 & 35,337 & 5,170 & 14.63 \\ 
        A24 & Tobacco; cigars; cigarettes; smokers' requisites & 67,262 & 3,230 & 360 & 11.15 \\ 
        A41 & Wearing apparel & 150,247 & 2,535 & 477 & 18.82 \\ 
        A42 & Headwear & 25,378 & 604 & 132 & 21.85 \\ 
        A43 & Footwear & 74,228 & 1,211 & 134 & 11.07 \\ 
        A44 & Haberdashery; jewellery & 48,613 & 864 & 72 & 8.33 \\ 
        A45 & Hand or travelling articles & 172,141 & 5,337 & 1,485 & 27.82 \\ 
        A46 & Brushware & 32,601 & 479 & 79 & 16.49 \\ 
        A47 & Furniture;domestic articles or appliances; coffee mills; spice mills;suction cleaners in general & 962,723 & 34,023 & 5,318 & 15.63 \\ 
        A61 & Medical or veterinary science; hygiene & 1,953,512 & 99,157 & 11,076 & 11.17 \\ 
        A62 & Life-saving; fire-fighting & 116,193 & 12,320 & 1,783 & 14.47 \\ 
        A63 & Sports; games; amusements & 376,930 & 11,391 & 740 & 6.5 \\ 
        A99 & Subject matter not otherwise provided for in section a & 229 & 33 & 4 & 12.12 \\ 
        B01 & Physical or chemical processes or apparatus in general & 1,332,667 & 610,307 & 120,771 & 19.79 \\ 
        B02 & Crushing, pulverising, or disintegrating; preparatory treatment of grain for milling & 228,879 & 36,527 & 7,753 & 21.23 \\ 
        B03 & Separation of solid materials using liquids or using pneumatic tables or jigs & 68,915 & 29,006 & 5,558 & 19.16 \\ 
        B04 & Centrifugal apparatus or machines for carrying-out physical or chemical processes & 28,461 & 4,358 & 624 & 14.32 \\ 
        B05 & Spraying or atomising in general; applying liquids or other fluent materials to surfaces, in general & 285,289 & 31,867 & 4,378 & 13.74 \\ 
        B06 & Generating or transmitting mechanical vibrations in general & 10,705 & 398 & 37 & 9.3 \\ 
        B07 & Separating solids from solids; sorting & 198,516 & 18,848 & 3,138 & 16.65 \\ 
        B08 & Cleaning & 437,757 & 71,364 & 16,123 & 22.59 \\ 
        B09 & Disposal of solid waste; reclamation of contaminated soil & 57,356 & 55,809 & 16,145 & 28.93 \\ 
        B21 & Mechanical metal-working without essentially removing material; punching metal & 426,389 & 16,896 & 1,196 & 7.08 \\ 
        B22 & Casting; powder metallurgy & 191,829 & 23,485 & 1,319 & 5.62 \\ 
        B23 & Machine tools; metal-working not otherwise provided for & 934,950 & 47,290 & 3,723 & 7.87 \\ 
        B24 & Grinding; polishing & 272,113 & 14,671 & 1,408 & 9.6 \\ 
        B25 & Hand tools; portable power-driven tools; tool handles & 435,173 & 15,788 & 1,096 & 6.94 \\ 
        B26 & Hand cutting tools; cutting; severing & 198,229 & 10,062 & 1,274 & 12.66 \\ 
        B27 & Working or preserving wood or similar material; nailing or stapling machines in general & 83,075 & 3,921 & 522 & 13.31 \\ 
        B28 & Working cement, clay, or stone & 171,793 & 12,839 & 1,735 & 13.51 \\ 
        B29 & Working of plastics; working of substances in a plastic state in general & 541,380 & 52,211 & 8,806 & 16.87 \\ 
        B30 & Presses & 67,870 & 11,163 & 2,079 & 18.62 \\ 
        B31 & Making paper articles; working paper & 54,618 & 1,712 & 230 & 13.43 \\ 
        B32 & Layered products & 271,991 & 29,462 & 4,588 & 15.57 \\ 
        B33 & Additive manufacturing technology & 58,020 & 14,563 & 436 & 2.99 \\ 
        B41 & Printing; lining machines; typewriters; stamps & 212,727 & 7,488 & 1,208 & 16.13 \\ 
        B42 & Bookbinding; albums; files; special printed matter & 47,299 & 870 & 61 & 7.01 \\ 
        B43 & Writing or drawing implements; bureau accessories & 69,639 & 615 & 92 & 14.96 \\ 
        B44 & Decorative arts & 52,726 & 1,840 & 312 & 16.96 \\ 
        B60 & Vehicles in general & 952,725 & 189,225 & 53,725 & 28.39 \\ 
        B61 & Railways & 93,671 & 9,141 & 1,076 & 11.77 \\ 
        B62 & Land vehicles for travelling otherwise than on rails & 396,508 & 27,174 & 5,156 & 18.97 \\ 
        B63 & Ships or other waterborne vessels; related equipment & 129,667 & 32,265 & 5,760 & 17.85 \\ 
\hline
    \end{tabular}}
\begin{tablenotes}
\footnotesize 
\item Note: The table reports patents, green patents, and text-verified green patents distribution across 3-digit CPC classification, together with the \% ratio of text-verified green over conventional green in the period 2010-2022 (part I). 
\end{tablenotes}
\end{table}
\end{landscape}

\begin{landscape}
\begin{table}[htbp]
    \centering
    \caption{Patents distribution by CPC (part II).}
        \label{tab: pat_dist_2}
\resizebox{1.3\textwidth}{!}{%
            \begin{tabular}{clcccc}
            \hline
        \textbf{Classification code} & \textbf{Description} & \textbf{Patents} & \textbf{Green patents} & \textbf{text-verified green patents} & \textbf{text-verified green\%} \\ 
        \hline
        B64 & Aircraft; aviation; cosmonautics & 151,284 & 23,935 & 3,208 & 13.4 \\ 
        B65 & Conveying; packing; storing; handling thin or filamentary material & 1,509,862 & 135,778 & 17,108 & 12.6 \\ 
        B66 & Hoisting; lifting; hauling & 340,304 & 13,225 & 1,425 & 10.78 \\ 
        B67 & Opening or closing bottles, jars or similar containers; liquid handling & 65,340 & 2,486 & 206 & 8.29 \\ 
        B68 & Saddlery; upholstery & 4,397 & 82 & 7 & 8.54 \\ 
        B81 & Microstructural devices or systems, e.g. micromechanical devices & 18,449 & 946 & 39 & 4.12 \\ 
        B82 & Nanotechnology & 71,531 & 19,314 & 3,051 & 15.8 \\ 
        B99 & Subject matter not otherwise provided for in section b & 5 & 1 & 1 & 100 \\ 
        C01 & Inorganic chemistry & 186,106 & 76,276 & 16,567 & 21.72 \\ 
        C02 & Treatment of water, waste water, sewage, or sludge & 424,569 & 424,558 & 119,694 & 28.19 \\ 
        C03 & Glass; mineral or slag wool & 84,510 & 15,840 & 1,973 & 12.46 \\ 
        C04 & Cements; concrete; artificial stone; ceramics; refractories & 123,836 & 37,926 & 7,804 & 20.58 \\ 
        C05 & Fertilisers; manufacture thereof & 43,954 & 30,734 & 7,160 & 23.3 \\ 
        C06 & Explosive; matches & 7,240 & 405 & 40 & 9.88 \\ 
        C07 & Organic chemistry & 394,040 & 99,820 & 11,371 & 11.39 \\ 
        C08 & Organic macromolecular compounds; their preparation or chemical working-up & 383,625 & 54,682 & 10,154 & 18.57 \\ 
        C09 & Dyes; paints, polishes, natural resins, adhesives; compositions not otherwise provided for & 288,443 & 49,330 & 6,766 & 13.72 \\ 
        C10 & Petroleum, gas or coke industries, technical gases containing carbon monoxide; fuels; lubricant; peat & 121,248 & 50,506 & 17,055 & 33.77 \\ 
        C11 & Animal or vegetable oils, fats, fatty substances or waxes; fatty acids therefrom; detergents; candles & 49,336 & 8,744 & 2,341 & 26.77 \\ 
        C12 & Biochemistry; beer; spirits; wine; vinegar; microbiology; enzymology; mutation or genetic engineering & 369,128 & 85,351 & 10,731 & 12.57 \\ 
        C13 & Sugar industry & 4,212 & 942 & 467 & 49.58 \\ 
        C14 & Skins; hides; pelts; leather & 9,732 & 422 & 74 & 17.54 \\ 
        C21 & Metallurgy of iron & 120,910 & 28,231 & 3,486 & 12.35 \\ 
        C22 & Metallurgy; ferrous or non-ferrous alloys; treatment of alloys or non-ferrous metals & 144,029 & 43,131 & 7,541 & 17.48 \\ 
        C23 & Coating metallic material; coating material with metallic material; chemical surface treatment & 159,300 & 21,953 & 2,770 & 12.62 \\ 
        C25 & Electrolytic or electrophoretic processes; apparatus therefor & 92,938 & 25,917 & 4,617 & 17.81 \\ 
        C30 & Crystal growth & 31,669 & 12,000 & 1,213 & 10.11 \\ 
        C40 & Combinatorial chemistry; libraries, e.g. chemical libraries, dna libraries & 4,883 & 570 & 20 & 3.51 \\ 
        C99 & Subject matter not otherwise provided for in section c & 6 & 3 & 0 & 0 \\ 
        D01 & Natural or man-made threads or fibres; spinning & 84,326 & 6,582 & 1,126 & 17.11 \\ 
        D02 & Yarns; mechanical finishing of yarns or ropes; warping or beaming & 26,964 & 1,039 & 133 & 12.8 \\ 
        D03 & Weaving & 40,326 & 1,176 & 131 & 11.14 \\ 
        D04 & Braiding; lace-making; knitting; trimmings; non-woven fabrics & 58,316 & 3,605 & 393 & 10.9 \\ 
        D05 & Sewing; embroidering; tufting & 42,533 & 548 & 22 & 4.01 \\ 
        D06 & Treatment of textiles or the like; laundering; flexible materials not otherwise provided for & 209,554 & 14,122 & 2,191 & 15.51 \\ 
        D07 & Ropes; cables other than electric & 8,945 & 200 & 16 & 8 \\ 
        D10 & Indexing scheme associated with sublasses of section d, relating to textiles & 15,129 & 1,199 & 164 & 13.68 \\ 
        D21 & Paper-making; production of cellulose & 51,382 & 7,804 & 2,050 & 26.27 \\ 
        D99 & Subject matter not otherwise provided for in section d & 9 & 1 & 1 & 100 \\ 
        E01 & Construction of roads, railways, or bridges & 323,643 & 34,608 & 6,812 & 19.68 \\ 
        E02 & Hydraulic engineering; foundations; soil shifting & 334,899 & 67,676 & 9,971 & 14.73 \\ 
        E03 & Water supply; sewerage & 193,719 & 116,768 & 18,110 & 15.51 \\ 
        E04 & Building & 798,477 & 168,343 & 29,548 & 17.55 \\ 
        E05 & Locks; keys; window or door fittings; safes & 230,644 & 7,442 & 966 & 12.98 \\ 
        E06 & Doors, windows, shutters, or roller blinds, in general; ladders & 205,466 & 21,419 & 4,070 & 19 \\ 
        E21 & Earth drilling; mining & 343,772 & 29,461 & 2,757 & 9.36 \\ 
 \hline \hline
    \end{tabular}}
\begin{tablenotes}
\footnotesize 
\item Note: The table reports patents, green patents, and text-verified green patents distribution across 3-digits CPC classification, together with the \% ratio of text-verified green over green - period 2010-2022 (part II). 
\end{tablenotes}
\end{table}
\end{landscape}

\begin{landscape}
\begin{table}[htbp]
    \centering
    \caption{Patents distribution by CPC (part III).}
        \label{tab: pat_dist_3}
\resizebox{1.3\textwidth}{!}{%
            \begin{tabular}{clcccc}
            \hline
        \textbf{Classification code} & \textbf{Description} & \textbf{Patents} & \textbf{Green patents} & \textbf{text-verified green patents} & \textbf{text-verified green\%} \\ 
        \hline
        E99 & Subject matter not otherwise provided for in section E & 10 & 0 & 0 & 0 \\ 
        F01 & Machines or engines in general; engine plants in general; steam engines & 194,068 & 91,189 & 10,496 & 11.51 \\ 
        F02 & Internal-combustion engines; hot-gas or combustion-product engine plants & 222,402 & 119,576 & 10,353 & 8.66 \\ 
        F03 & Machines or engines for liquids; wind, spring, weight, or miscellaneous motors & 102,544 & 88,059 & 18,877 & 21.44 \\ 
        F04 & Engines or pumps & 317,546 & 27,282 & 4,048 & 14.84 \\ 
        F05 & Indexing schemes relating to engines or pumps in various subclasses of classes f01-f04 & 61,699 & 35,014 & 2,689 & 7.68 \\ 
        F15 & Fluid-pressure actuators; hydraulics or pneumatics in general & 86,451 & 4,551 & 536 & 11.78 \\ 
        F16 & Engineering elements or units; general measures for effecting and maintaining effective functioning of machines & 1,413,801 & 105,884 & 13,355 & 12.61 \\ 
        F17 & Storing or distributing gases or liquids & 71,961 & 11,870 & 931 & 7.84 \\ 
        F21 & Lighting & 482,428 & 84,562 & 29,276 & 34.62 \\ 
        F22 & Steam generation & 41,352 & 14,987 & 5,195 & 34.66 \\ 
        F23 & Combustion apparatus; combustion processes & 157,212 & 81,683 & 20,375 & 24.94 \\ 
        F24 & Heating; ranges; ventilating & 551,508 & 198,801 & 76,181 & 38.32 \\ 
        F25 & Refrigeration or cooling; combined heating and refrigeration systems; heat pump systems & 210,520 & 49,008 & 12,255 & 25.01 \\ 
        F26 & Drying & 201,694 & 27,839 & 7,292 & 26.19 \\ 
        F27 & Furnaces; kilns; ovens; retorts & 103,068 & 32,364 & 9,155 & 28.29 \\ 
        F28 & Heat exchange in general & 143,847 & 39,355 & 9,325 & 23.69 \\ 
        F41 & Weapons & 52,780 & 1,088 & 65 & 5.97 \\ 
        F42 & Ammunition; blasting & 34,595 & 1,260 & 74 & 5.87 \\ 
        F99 & Subject matter not otherwise provided for in section f & 4 & 0 & 0 & ~ \\ 
        G01 & Measuring; testing & 2,344,888 & 171,704 & 16,917 & 9.85 \\ 
        G02 & Optics & 482,153 & 22,468 & 1,966 & 8.75 \\ 
        G03 & Photography; cinematography; analogous techniques using waves other than optical waves & 246,789 & 7,373 & 541 & 7.34 \\ 
        G04 & Horology & 34,830 & 1,533 & 249 & 16.24 \\ 
        G05 & Controlling; regulating & 368,198 & 69,261 & 9,241 & 13.34 \\ 
        G06 & Computing; calculating; counting & 1,983,168 & 187,084 & 14,218 & 7.6 \\ 
        G07 & Checking-devices & 190,431 & 10,242 & 1,897 & 18.52 \\ 
        G08 & Signalling & 341,739 & 36,256 & 6,471 & 17.85 \\ 
        G09 & Educating; cryptography; display; advertising; seals & 464,858 & 27,671 & 4,451 & 16.09 \\ 
        G10 & Musical instruments; acoustics & 120,526 & 6,015 & 516 & 8.58 \\ 
        G11 & Information storage & 112,199 & 4,806 & 122 & 2.54 \\ 
        G12 & Instrument details & 4,565 & 179 & 17 & 9.5 \\ 
        G16 & Information and communication technology [ict] specially adapted for specific application fields & 85,329 & 8,602 & 451 & 5.24 \\ 
        G21 & Nuclear physics; nuclear engineering & 40,347 & 22,718 & 755 & 3.32 \\ 
        G99 & Subject matter not otherwise provided for in section g & 19 & 1 & 0 & 0 \\ 
        H01 & Electric elements & 1,961,055 & 610,482 & 84,510 & 13.84 \\ 
        H02 & Generation, conversion, or distribution of electric power & 1,234,321 & 436,827 & 156,797 & 35.89 \\ 
        H03 & Electronic circuitry & 175,601 & 11,017 & 441 & 4 \\ 
        H04 & Electric communication technique & 1,726,290 & 126,570 & 11,399 & 9.01 \\ 
        H05 & Electric techniques not otherwise provided for & 656,296 & 118,823 & 13,972 & 11.76 \\ 
        H10 & Semiconductor devices; electric solid-state devices not otherwise provided for & 301,283 & 120,771 & 27,454 & 22.73 \\ 
        H99 & Subject matter not otherwise provided for in section H & 24 & 3 & 1 & 33.33 \\ 
        Y02 & Technologies or applications for mitigation or adaptation against climate change & 2,442,727 & 2,439,432 & 489,298 & 20.06 \\ 
        Y04 & Information or communication technologies having an impact on other technology areas & 55,317 & 55,249 & 9,227 & 16.7 \\ 
        Y10 & Technical subjects covered by former uspc & 172,307 & 29,631 & 4,990 & 16.84 \\  \hline \hline
    \end{tabular}}
\begin{tablenotes}
\footnotesize 
\item Note: The table reports patents, green patents, and text-verified green patents distribution across 3-digits CPC classification, together with the \% ratio of text-verified green over green - period 2010-2022 (part III). 
\end{tablenotes}
\end{table}
\end{landscape}


\begin{table}[htbp]
\centering
\caption{Top 10 Patent Office and Technology Class pairs by Revealed Comparative Advantage (RCA)}
        \label{tab: rca_dist_1}
\resizebox{\textwidth}{!}{%
\begin{tabular}{llp{6.5cm}ccccc}
\toprule
\textbf{Patent office} & \textbf{Code} & \textbf{Description} & \textbf{Patents} & \textbf{\makecell[r]{Green\\patents}} & \textbf{\makecell[r]{Text-verified\\green patents}} & \textbf{\makecell[r]{RCA\\green}} & \textbf{\makecell[r]{RCA text-\\verified green}} \\
\midrule
\multicolumn{8}{l}{\textit{Top 10 text-verified green}} \\
\midrule
Greece & H10 & Semiconductor devices; electric solid-state devices not otherwise provided for & 38 & 35 & 31 & 17.46 & 44.16 \\
United Kingdom & D02 & Yarns; mechanical finishing of yarns or ropes; warping or beaming & 38 & 9 & 4 & 7.79 & 24.39 \\
Denmark & B68 & Saddlery; upholstery & 33 & 1 & 1 & 1.65 & 22.65 \\
Japan & C40 & Combinatorial chemistry; libraries, e.g. chemical libraries, dna libraries & 278 & 79 & 11 & 3.35 & 21.01 \\
Bulgaria & B66 & Hoisting; lifting; hauling & 41 & 3 & 3 & 1.95 & 18.78 \\
Brazil & G11 & Information storage & 51 & 3 & 1 & 1.40 & 18.46 \\
Portugal & A63 & Sports; games; amusements & 102 & 7 & 3 & 2.36 & 15.41 \\
Slovakia & F41 & Weapons & 57 & 2 & 1 & 1.73 & 14.67 \\
Serbia & A63 & Sports; games; amusements & 40 & 2 & 1 & 1.69 & 13.01 \\
Brazil & D02 & Yarns; mechanical finishing of yarns or ropes; warping or beaming & 34 & 2 & 2 & 1.56 & 12.76 \\
\multicolumn{8}{l}{\textit{Top 10 green}} \\
\midrule
Russia & Y02 & Technologies or applications for mitigation or adaptation against climate change & 19,783 & 19,782 & 3,616 & 26.92 & 0.89 \\
Finland & C11 & Animal or vegetable oils, fats, fatty substances or waxes; fatty acids therefrom; detergents; candles & 67 & 54 & 10 & 19.39 & 3.52 \\
South Africa & H10 & Semiconductor devices; electric solid-state devices not otherwise provided for & 52 & 48 & 20 & 17.96 & 6.23 \\
Denmark & F05 & Indexing schemes relating to engines or pumps in various subclasses of classes f01-f04 & 1,159 & 1,110 & 39 & 17.79 & 0.76 \\
Greece & H10 & Semiconductor devices; electric solid-state devices not otherwise provided for & 38 & 35 & 31 & 17.46 & 44.16 \\
Canada & C13 & Sugar industry & 84 & 69 & 35 & 17.15 & 6.10 \\
Belgium & B22 & Casting; powder metallurgy & 40 & 28 & 1 & 16.73 & 3.70 \\
Denmark & B66 & Hoisting; lifting; hauling & 335 & 106 & 1 & 11.52 & 0.71 \\
Poland & B09 & Disposal of solid waste; reclamation of contaminated soil & 410 & 409 & 74 & 11.43 & 0.56 \\
Switzerland & F28 & Heat exchange in general & 56 & 45 & 8 & 10.86 & 2.40 \\
\bottomrule
\end{tabular}%
}
\begin{tablenotes}
\footnotesize 
\item Note: The table reports top 10 country-CPC pairs by RCA index separately for green patents and text-verified green patents - period 2010-2022. 
\end{tablenotes}
\end{table}

\end{document}